\documentclass[twocolumn, times]{aastex61}

\usepackage{graphicx, epsfig, amssymb} 
\usepackage{amsmath, amsfonts}
\usepackage{bm} 
\usepackage{pstricks,psfrag,pst-node,pst-text,pst-3d,pst-grad}

\newcommand{\dif}{{\rm d}}
\newcommand{\del}{\partial}
\newcommand{\urr}{u_{r}}

\newcommand{\uth}{u_{\theta}}
\newcommand{\uph}{u_{\varphi}}
\newcommand{\eps}{\varepsilon}
\newcommand{\upar}{u_{\parallel}}
\newcommand{\uper}{u_{\perp}}

\newcommand{\Br}{B_{r}}
\newcommand{\Bth}{B_{\theta}}
\newcommand{\Bph}{B_{\varphi}}

\newcommand{\rs}{r_s}
\newcommand{\rnu}{r_{\nu}}

\newcommand{\Pm}{\mathcal{P}}

\newcommand{\mnab}{\bm{\nabla}}
\newcommand{\mvel}{\bm{u}}

\newcommand{\mB}{\bm{B}}
\newcommand{\mE}{\bm{E}}

\newcommand{\mn}{\bm{n}}

\newcommand{\Bpar}{B_{\parallel}}
\newcommand{\Bper}{B_{\perp}}

\received{--}
\revised{--}
\accepted{--}
\submitjournal{ApJ}

\shorttitle{Rotation and magnetic field in supernova explosion}
\shortauthors{Fujisawa, Okawa et al.}

\begin{document}

\title{Effects of rotation and magnetic field on the revival of a stalled shock in supernova explosions}

\correspondingauthor{Kotaro Fujisawa, Hirotada Okawa. These two authors contributed equally to this work.}
\email{fujisawa@heap.phys.waseda.ac.jp, h.okawa@aoni.waseda.jp}


\author{Kotaro Fujisawa}
\affil{Advanced Research Institute for Science and Engineering, 
Waseda University, 3-4-1 Okubo, Shinjuku-ku, 
Tokyo 169-8555, Japan \\}

\author{Hirotada Okawa}
\affil{Yukawa Institute for Theoretical Physics, Kyoto University, Kyoto 606-8502, Japan}
\affil{Advanced Research Institute for Science and Engineering, 
Waseda University, 3-4-1 Okubo, Shinjuku-ku, 
Tokyo 169-8555, Japan \\}

\author{Yu Yamamoto}
\affil{Advanced Research Institute for Science and Engineering, 
Waseda University, 3-4-1 Okubo, Shinjuku-ku, 
Tokyo 169-8555, Japan \\}

\author{Shoichi Yamada}
\affil{Advanced Research Institute for Science and Engineering, 
Waseda University, 3-4-1 Okubo, Shinjuku-ku, 
Tokyo 169-8555, Japan \\}
\affil{Science \& Engineering, Waseda University, 3-4-1 Okubo, Shinjuku-ku, 
Tokyo 169-8555, Japan \\}



 \begin{abstract}
  We investigate axisymmetric steady solutions of (magneto)hydrodynamics equations that describe
  approximately accretion flows through a standing shock wave
  and discuss the effects of rotation and magnetic field on the revival of the stalled shock wave
  in supernova explosions.
  We develop a new powerful numerical method to calculate
  the 2-dimensional (2D) steady accretion flows self-consistently. 
  We first confirm the results of preceding papers that there is  a critical luminosity
  of irradiating neutrinos, above which there exists no steady solution in spherical models.
  If a collapsing star has rotation and/or magnetic field, the accretion flows
  are no longer spherical owing to the centrifugal force and/or Lorentz force
  and the critical luminosity is modified.
  In fact we find that the critical luminosity is reduced by about $50\%$ -- $70\%$ for rapid rotations
  and about $20\%$ -- $50\%$ for strong toroidal magnetic fields,
  depending on the mass accretion rate.
  These results may be also interpreted as an existence of the critical specific angular momentum
  or critical magnetic field, above which there exists no steady solution and
  the standing shock wave will revive for a
  given combination of mass accretion rate and neutrino luminosity. 
 \end{abstract}






\section{Introduction}\label{sec:intro}
%
Core-collapse supernovae (CCSNe) play important roles in various fields of astrophysics
such as the star formation, galactic evolution and the acceleration of
cosmic-ray through  their nucleosynthesis, energetic shock waves and high
luminosities. The large gravitational energy liberated in the collapse makes
CCSNe promising sites for the emissions of neutrinos and gravitational waves
as well as heavy elements. The understanding of the physical processes in
and the mechanism of CCSNe is important for the development of 
multi-messenger astronomy.

CCSNe commence with the gravitational collapse of the core of massive stars at the
ends of their lives.  When the central density reaches the nuclear
saturation density, core bounce occurs and a shock wave is generated.
As it propagates outward, 
the shock loses energy via photo-dissociations of nuclei and neutrino cooling
and is eventually stagnated inside the core by the ram pressure
of accreting matter in addition to these energy losses.
The stalled shock wave should be revived somehow to produce a successful explosion.
It is widely expected that it will be re-energized by the irradiation
of matter behind it by neutrinos diffusing out of a proto-neutron star (PNS)
(\citealp{Wilson:1985}).
This neutrino heating scenario is currently the most favored mechanism of CCSNe 
(\citealp{Janka_et_al:2012}; \citealp{Kotake_et_al:2012}; \citealp{Burrows:2013};
\citealp{Foglizzo_et_al:2015}; \citealp{Muller:2016}; \citealp{Burrows_et_al:2018} for recent reviews).

Intensive and extensive efforts in numerical simulations have revealed so far
that no successful explosion obtains in spherical symmetry 
(e.g.~\citealp{Liebendorfer_et_al:2001,Liebendorfer_et_al:2005};
\citealp{Rampp_Janka:2002}; \citealp{Sumiyoshi_et_al:2005})
and multi-dimensional effects are crucially important 
(\citealt{Burrows_et_al:2006}; \citealt{Marek_Janka:2009}; \citealt{Bruenn_et_al:2009};
\citealt{Suwa_et_al:2010}; \citealt{Muller_Janka_Marek:2012}; \citealt{Takiwaki_et_al:2012};
\citealt{Hanke_et_al:2013}; \citealt{Murphy_Dolence_Burrows:2013}; \citealt{Couch:2013};
\citealt{Couch_Ott:2013}; \citealt{Lentz_et_al:2015};
\citealt{Nakamura_et_al:2015}; \citealt{Melson_et_al:2015};
\citealt{Roberts_et_al:2016}; \citealt{Bruenn_et_al:2016}; \citealt{OConnor_Couch:2018}). 
Important effects among them are rotation 
(\citealt{Fryer_Heger:2000}; \citealt{Kotake_et_al:2003}; \citealt{Thompson_et_al:2005};
\citealt{Marek_Janka:2009}; \citealt{Iwakami_Nagakura_Yamada:2014b};
\citealt{Nakamura_et_al:2014}; \citealt{Takiwaki_et_al:2016}; \citealt{Summa_et_al:2018}),
magnetic field
(\citealt{Akiyama_et_al:2003}; \citealt{Yamada_Sawai:2004}; \citealt{Kotake_et_al:2004};
\citealt{Sawai_Kotake_Yamada:2005}; \citealt{Obergaulinger_et_al:2006};
\citealt{Burrows_et_al:2007};
\citealt{Takiwaki_Kotake_Sato:2009}; \citealt{Sawai_Yamada:2014};
\citealt{Obergaulinger_et_al:2014}; \citealt{Mosta_et_al:2015}; \citealt{Sawai_Yamada:2016};
\citealt{Obergaulinger_et_al:2018}),
non-spherical structures of the progenitor
(\citealt{Couch_Ott:2013}; \citealt{Takahashi_Yamada:2014}; \citealt{Couch_et_al:2015}; \citealt{Takahashi_et_al:2016}),
turbulence 
(\citealt{Murphy_Burrows:2008}; \citealt{Murphy_Meakin:2011};
\citealt{Murphy_Dolence_Burrows:2013}; \citealt{Couch_Ott:2015}; \citealt{Mabanta_Murphy:2018}),
(magneto)hyrdodynamical instabilities  
(\citealt{Blondin_et_al:2003}; \citealt{Scheck_et_al:2006};
\citealt{Blondin_Mezzacappa:2007};
\citealt{Iwakami_et_al:2008}; \citealt{Guilet_et_al:2010};
\citealt{Wongwathanarat_et_al:2010};
\citealt{Takiwaki_et_al:2014}; \citealt{Fernandez_et_al:2014}; \citealt{Fernandez:2015}),
general relativistic gravity 
(\citealt{Dimmelmeier_et_al:2002}; \citealt{Shibata_Sekiguchi:2004, Shibata_Sekiguchi:2005},
\citealt{Muller_Janka_Marek:2012}; \citealt{Ott_et_al:2012}; \citealt{Kuroda_et_al:2012,Kuroda_et_al:2016}) 
and neutrino transport 
(\citealt{Nagakura_et_al:2014}; \citealt{Dolence_Burrows_Zhang:2015};
\citealt{Pan_et_al:2016}; \citealt{Nagakura_et_al:2017, Nagakura_et_al:2018}).
It is true that large-scale dynamical simulations have played a crucial role in the recent
progresses in our understanding of these ingredients, but we believe
that a more phenomenological approach that employs toy models 
still plays an indispensable and
complementary role to understand each effect more deeply. 

\citet{Burrows_Goshy:1993} took such an approach and
introduce the concept of the critical neutrino luminosity. 
They approximated the accretion flows after the shock wave is stagnated
with spherical steady solutions with constant mass accretion rates of the
hydrodynamics (HD) equations that incorporate neutrino heating of matter; 
they found then that for a given mass accretion rate,
there is  a critical neutrino luminosity, above which no steady solution exists;
they surmised that the revival of the stalled shock wave should occur there.
The existence of such a critical neutrino luminosity was confirmed both
in similar phenomenological methods 
(\citealp{Yamasaki_Yamada:2005, Yamasaki_Yamada:2007}; \citealt{Keshet_Balberg:2012};
\citealt{Murphy_Dolence:2017})
and in numerical simulations (\citealp{Janka_Mueller:1996}; \citealt{Ohnishi_Kotake_Yamada:2006};
\citealt{Iwakami_et_al:2008}; \citealt{Murphy_Burrows:2008}; \citealt{Nordhaus_et_al:2010};
\citealt{Iwakami_Nagakura_Yamada:2014a,Iwakami_Nagakura_Yamada:2014b}; \citealt{Suwa_et_al:2016}).

For examples in the phenomenological models, 
\citet{Pejcha_Thompson:2012} contended that the ante-sonic condition
for the ratio of the sound velocity to the
escape velocity is an important diagnostic for explosion.
(see also \citealt{Raives_et_al:2018}). 
\citet{Nagakura_Yamamoto_Yamada:2013} argued, on the other hand, 
that there is a critical fluctuation for a given accretion flow, above
which the shock wave can sustain its outward propagation once launched.
\citet{Gabay_et_al:2015} applied a phase-space analysis to
the shock acceleration in the shock radius and shock velocity plane.
More recently \citet{Murphy_Dolence:2017} contended that five parameters
(mass accretion rate, neutrino luminosity, neutrino temperature,
PNS mass and radius) instead of two (mass accretion rate and
neutrino luminosity alone) are more appropriate in considering the condition
for explosion. They also derived a diagnostic dimensionless quantity $\Psi$ from
the integrability condition, which roughly measures the balance between pressure and gravity.
It should be stressed here that all these conditions are based on the analysis
of spherical accretion flows and should 
be  modified if the background flow is non-spherical.

In fact, it was shown that the critical luminosity is
lowered by rotation both in the phenomenological method (\citealp{Yamasaki_Yamada:2005})
and dynamical simulations (\citealp{Murphy_Burrows:2008};
\citealt{Nordhaus_et_al:2010}; \citealt{Hanke_et_al:2013}; \citealt{Couch:2013};
\citealp{Iwakami_Nagakura_Yamada:2014a,Iwakami_Nagakura_Yamada:2014b}).
\citet{Yamasaki_Yamada:2005} extended their spherically-symmetric, steady accretion-flow
to axisymmetric ones, taking rotation into account consistently.
They showed that the accretion flows become non-radial in the presence of rotation
and the critical luminosity is lowered.
\citet{Iwakami_Nagakura_Yamada:2014b}, on the other hand,
performed 3-dimensional (3D) dynamical simulations of rotational accretions 
and obtained similar results: the critical luminosity is lowered by rapid rotation
and there is a critical value of specific angular momentum,
above which the shock wave is revived, for a given combination of
mass accretion rate and neutrino luminosity.
This implies that an explosion may be triggered by rapid rotation even
if the neutrino luminosity is sub-critical without rotation.

It is emphasized here that \citet{Yamasaki_Yamada:2005} is the only one
that has so far taken the phenomenological approach to the non-spherical accretion flows.
This is in sharp contrast to the fact that there have been many numerical simulations
that are meant to address the issue (\citealp{Janka_Mueller:1996}; \citealt{Ohnishi_Kotake_Yamada:2006};
\citealt{Iwakami_et_al:2008}; \citealt{Murphy_Burrows:2008}; \citealt{Nordhaus_et_al:2010};
\citealt{Hanke_et_al:2013};
\citealt{Iwakami_Nagakura_Yamada:2014a,Iwakami_Nagakura_Yamada:2014b}; \citealt{Suwa_et_al:2016}).
The reason for this is probably the difficulty to obtain numerically non-spherical steady
solutions of the HD equations. Even in \citet{Yamasaki_Yamada:2005},
the number of solutions was not large. Not to mentions, there has been no
attempt to incorporate magnetic field in such approaches. 

In this paper, we develop a new numerical scheme to numerically obtain axisymmetric steady
solutions of HD and magnetohydrodynamics (MHD) equations that describe self-consistently
 non-spherical post-shock accretion flows in the presence of rotation and/or magnetic field.
 Based on these solutions we  study the effects of rotation and
 magnetic field on the critical neutrino luminosity systematically.
 This paper is organized as follows. In Sec.~\ref{sec:basic} we formulate
 the problem and give the basic equations to be solved.
 We then explain our new numerical method.
 In Sec.~\ref{sec:results} we present numerical solutions and the critical luminosities
 obtained from them.  Finally we give some discussions and summarize this paper in Sec.~\ref{sec:discussion}.

\section{Formulations}\label{sec:basic}

\subsection{Assumptions and basic equations}

After the shock stalls, the mass accretion rate and neutrino luminosity change slowly
and the post-shock accretion flows may be approximated as steady states with
constant mass accretion rates and neutrino luminosities
(\citealt{Burrows_Goshy:1993}; \citealt{Yamasaki_Yamada:2005}).
We make the following assumptions to construct idealized models in this paper. 
\begin{enumerate}
 \item  The equatorial symmetry is assumed in addition to the stationary 
	($\partial / \partial t = 0$) and axisymmetry ($\partial / \partial \varphi = 0$).

 \item We ignore both viscosity and magnetic resistivity and consider only ideal HD
       or MHD equations. 

 \item We consider the accretion flows only in the domain between the
       stalled shock wave and the neutrino sphere and impose
       the outer and inner boundary conditions there. 
       
 \item The PNS is treated as a point mass with $1.3~\mathrm{M_\odot}$.
       Only its Newtonian gravitational forces are considered and the self-gravity
       of accreting matter is ignored for simplicity. 
       
 \item We neglect the neutrino transfer again for simplicity
       and the luminosity and energy spectrum of neutrinos are
       assumed to be constant in radius (see Equations~\ref{eq:qdot} and~\ref{eq:L_nu}).

 \item A simplified equation of state (EOS) that neglects the 
       photo-dissociations of nuclei is adopted and the convection 
       is not taken into account. 
\end{enumerate}
Under these assumptions,
the basic equations are given as 
\begin{subequations}
 \label{eq:basics_hydro}
  \begin{align}
   \label{eq:continuity}
   \frac{1}{r^2}&\frac{\del}{\del r}\left(r^2\rho \urr\right) 
   +\frac{1}{r\sin\theta}\frac{\del}{\del \theta}\left(\sin\theta \rho u_{\theta}\right) = 0,
   \\
   \label{eq:momentum_r}
    \urr&\frac{\del \urr}{\del r} +\frac{\uth}{r}\frac{\del \urr}{\del \theta} 
    -\frac{\uth^2+\uph^2}{r} = -\frac{1}{\rho}\frac{\del p}{\del r}-\frac{GM}{r^2}
   \nonumber \\ 
   &-\frac{1}{4\pi\rho}
    \Bigg[
    \Bth\frac{\del\Bth}{\del r}
    +\Bph\frac{\del\Bph}{\del r}
    +\frac{\Bth^2+\Bph^2}{r}
    \Bigg],
    \\
    \label{eq:momentum_th}
    \urr&\frac{\del \uth}{\del r} +\frac{\uth}{r}\frac{\del \uth}{\del \theta}
    +\frac{\urr\uth}{r} -\frac{\uph^2\cot\theta}{r}
    = -\frac{1}{\rho r}\frac{\del p}{\del\theta}
   \nonumber \\
   &+\frac{1}{4\pi\rho}
    \Bigg[
    \Br\frac{\del\Bth}{\del r}
    -\frac{\Br}{r}\frac{\del\Br}{\del\theta}
   \nonumber \\
   &-\frac{\Bph}{r}\frac{\del\Bph}{\del\theta}
    +\frac{\Br\Bth -\Bph^2\cot\theta}{r}
    \Bigg]
    ,\\
    \label{eq:momentum_ph}
    \urr&\frac{\del \uph}{\del r} +\frac{\uth}{r}\frac{\del \uph}{\del \theta}
   +\frac{\uph\urr}{r} +\frac{\uth\uph\cot\theta}{r}
   \nonumber \\
    &= \frac{1}{4\pi\rho}
    \left[
    \Br\frac{\del\Bph}{\del r}
    +\frac{\Bth}{r}\frac{\del\Bph}{\del\theta}
    +\frac{\Br\Bph -\Bth\Bph\cot\theta}{r}
    \right],
    \\
    \label{eq:energy}
    \urr&\left(\frac{\del \eps}{\del r} -\frac{p}{\rho^2}\frac{\del \rho}{\del r}\right)
    +\frac{\uth}{r}\left(\frac{\del \eps}{\del\theta}
    -\frac{p}{\rho^2}\frac{\del \rho}{\del\theta}\right) = \dot{q},
  \end{align}
\end{subequations}
where $\rho, u_i, B_i, p, \eps$ and $\dot{q}$ denote, respectively,
the density, velocity, magnetic field, pressure, specific internal energy and neutrino-heating rate;
$G$ and $M$ are the gravitational constant and the mass of PNS.
In the absence of the magnetic field, $B$ is simply set to zero.
In its presence, on the other hand, we also solve the Maxwell equations under the conditions of stationary 
and ideal MHD (i.e. $\mnab\cdot\mB = 0,\quad \mnab\times \mE = 0,\quad\mE=-\mvel\times\mB$)
as follows:
\begin{subequations}
 \label{eq:basics_mag}
  \begin{align}
   \label{eq:mag_con}
   \frac{1}{r^2}&\frac{\del}{\del r}\left(r^2 \Br\right) 
   +\frac{1}{r\sin\theta}\frac{\del}{\del \theta}\left(\sin\theta \Bth\right) = 0, \\
   \label{eq:mag_field_r}
   -&\uth\frac{\partial\Br}{\partial \theta}
   +\urr\frac{\partial\Bth}{\partial \theta}
   -\Br\frac{\partial\uth}{\partial \theta}
   \nonumber \\
   &+\Bth\frac{\partial\urr}{\partial \theta}
   +\urr\Bth\cot\theta
   -\uth\Br\cot\theta
   = 0,\\
   \label{eq:mag_field_th}
   r&\uth\frac{\partial\Br}{\partial r}
   -r\urr\frac{\partial\Bth}{\partial r}
   -r\Bth\frac{\partial\urr}{\partial r}
   \nonumber \\
   &+r\Br\frac{\partial\uth}{\partial r}
   +\Br\uth -\Bth\urr
   = 0,\\
   \label{eq:mag_field_ph}
   r&\Br\frac{\partial\uph}{\partial r}
   -r\Bph\frac{\partial\urr}{\partial r}
   +r\uph\frac{\partial\Br}{\partial r}
   -r\urr\frac{\partial\Bph}{\partial r}
   \nonumber \\
   &+\Bth\frac{\partial\uph}{\partial \theta}
   -\Bph\frac{\partial\uth}{\partial \theta}
   +\uph\frac{\partial\Bth}{\partial \theta}
   \nonumber \\
   &-\uth\frac{\partial\Bph}{\partial \theta}
   +\Br\uph -\Bph\urr
   = 0,
   \end{align}
\end{subequations}
where $E_i$ is the electric field.
Following \citet{Yamasaki_Yamada:2005}, we employ a simplified 
EOS given as follows:
 \begin{subequations} 
\begin{align}
 p= \frac{11\pi^2}{180}\frac{k^4}{c^3\hbar^3}T^4 +\frac{\rho kT}{m_N}, \\
  \eps = \frac{11\pi^2}{60}\frac{k^4}{c^3\hbar^3}\frac{T^4}{\rho} +\frac{3}{2}\frac{kT}{m_N}, 
\end{align}
 \end{subequations}
where $k,c,\hbar$, $m_N$ and $T$ are the Boltzmann constant, speed of light,
reduced Planck constant, nucleon mass and matter temperature, respectively.
Note that this EOS neglects photo-dissociations of nuclei and, as a consequence,
the critical luminosity tends to be overestimated.
Since we are interested in the qualitative effects of rotation and/or magnetic field
on the critical luminosity in this paper, the simplification is justified.
The net neutrino-heating rate is also simplified as in \citet{Herant:1992}
\begin{eqnarray}
 \label{eq:qdot}
 \dot q &=& 4.8\times
  10^{32}\left[1-\sqrt{1-\frac{r_{\nu}^2}{r^2}}\right]\frac{L_{\nu}}{2\pi r^2_{\nu}}T_{\nu}^2
   \nonumber \\
  &-&2.0\times 10^{18}T^6 \ \mathrm{\left[\mathrm{ergs}\ \mathrm{s}^{-1}\ \mathrm{g}^{-1}\right]},
\end{eqnarray}
where the neutrino luminosity $L_{\nu}$ and temperature $T_{\nu}$
are model parameters, which are constant both in time and space, and common to
$\nu_e$ and $\bar{\nu}_e$; we assume further that they are related with the 
radius of neutrino sphere~$r_{\nu}$, which is coincident with the inner boundary
in our models, as
\begin{eqnarray}
 \label{eq:L_nu}
 L_{\nu} = \frac{7}{4}\pi r_{\nu}^2\sigma_S T^4_{\nu},\label{eq:neutrino_sphere}
\end{eqnarray}
where $\sigma_S$ denotes the Stefan-Boltzmann constant.
The neutrino temperature, which is supposed to characterize the neutrino energy
spectrum, is set to $T_{\nu} = 4.5~{\mathrm{MeV}}$ in the following (\citealt{Yamasaki_Yamada:2005}). 

  \subsection{Boundary conditions}\label{ssec:shock_bc}

The region of our interest is the one between the neutrino sphere and the
stalled shock wave, and the inner and outer boundary conditions are imposed there, respectively.
Following \citet{Yamasaki_Yamada:2005}, we impose $\rho = 10^{11}\,\mathrm{g~cm^{-3}}$
on the inner boundary, which approximately corresponds to the condition that
the optical depth to the neutrino sphere from infinity is equal to $2/3$.
The latter condition was adopted by \citet{Burrows_Goshy:1993} and \cite{Murphy_Burrows:2008}
(see \citealt{Keshet_Balberg:2012} for a comparison of these two conditions).
The radius of the neutrino sphere is obtained from the neutrino luminosity and temperature
in Equation~(\ref{eq:L_nu}).

The main difficulty in obtaining steady accretion flows through
the stalled shock wave in multi-dimensions
stems from the simple fact that the shock surface
is non-spherical. In this paper, we assume that the shape of shock surface
expressed as $r=\rs(\theta)$ in the polar coordinates
is given with the Legendre function $\Pm_2$ as
\begin{subequations}
 \begin{align}
 \rs (\theta) =& r_{\mathrm{sph}}\left\{1 -\epsilon\Pm_2(\cos\theta)\right\},\\ 
 \Pm_2(\cos\theta) =& \frac{1}{2}\left(3\cos^2\theta-1\right),
 \end{align}
 \label{eq:rs}
\end{subequations}
where $r_{\mathrm{sph}}$ and $\epsilon$ are the equatorial radius
and deformation amplitude of the shock surface, respectively.
Instead of attempting to obtain the shock surface directly,
we employ these two degrees of freedom alone, which turn out to be sufficient.
(see Sec.~\ref{ssec:parameter}).
Once the shock shape is determined, the post-shock quantities are obtained
from the pre-shock quantities via the Rankine-Hugoniot relations,
which are given in Appendix~\ref{sec:shock}.
We assume that the pre-shock flows are spherically symmetric free falls from
infinity with different mass accretion rates. We do not attempt in this paper to
elaborate on the outer flow so that rotation and/or magnetic field would be taken into account
consistently.

Other boundary conditions are administered at
$\theta=0$ and $\theta=\pi/2$. As in \cite{Yamasaki_Yamada:2005},
we impose Neumann conditions for $\rho, \urr, T, \Br$ at $\theta = 0,\pi/2$, and
for $\uph, \Bph$ at $\theta = \pi/2$ while we set
Dirichlet conditions for $\uth$, $\Bth$, $\uph$, $\Bph$ at $\theta=0$ and
for $\uth$ and $\Bth$ also at $\theta=\pi/2$
as follows:
\begin{subequations}
 \label{eq:boundary_condition}
  \begin{align}
   \uth = \Bth = \uph = \Bph = 0
   \\
   \frac{\del \rho}{\del \theta} =  \frac{\del \urr}{\del \theta} =
  \frac{\del T}{\del \theta} = \frac{\del \Br}{\del \theta} =  0
  \end{align}
\end{subequations}
at $\theta = 0$ and
\begin{subequations}
 \label{eq:boundary_condition2}
  \begin{align}
   \uth = \Bth = 0
   \\
   \frac{\del \rho}{\del \theta} =  \frac{\del \urr}{\del \theta} =
   \frac{\del T}{\del \theta} = \frac{\del \Br}{\del \theta} =
   \frac{\del \uph}{\del \theta} =  \frac{\del \Bph}{\del \theta} = 0
  \end{align}
\end{subequations}
at $\theta = \pi / 2$ respectively. 

\subsection{Rotation law and magnetic field profiles}\label{ssec:rotation_law}

In the presence of magnetic field, the rotation law and the magnetic field profile
should be given consistently with each other, since
there are some conserved quantities along the streamline,
which constrain the variations of rotation and magnetic field, for the 
axisymmetric and steady ideal MHD flows
(see \citealt{Lovelace_et_al:1986}; \citealt{Fujisawa_et_al:2013}).
The details of their derivations are described in Appendix~\ref{sec:derivation}.

If a magnetic field is purely toroidal ($\Br = \Bth = 0$),
the specific angular momentum $\ell$ and  a quantity related with magnetic torque $\sigma$ are
conserved along the streamline:
\begin{subequations}
 \begin{align}
  \ell(\psi) = r \sin \theta \uph, \\
  \sigma(\psi) = \frac{\Bph}{r \sin \theta \rho},
 \end{align}
 \label{eq:ell_sigma} 
\end{subequations} 
where $\psi$ is a stream function defined as
 \begin{eqnarray}
  \urr \equiv \frac{1}{4 \pi \rho r^2 \sin \theta } \frac{\del \psi}{\del \theta}, \
   \uth \equiv -\frac{1}{4\pi \rho r \sin \theta} \frac{\del \psi}{\del r}.
  \label{eq:stream_function}
 \end{eqnarray}
Instead of fixing the functional forms of $\ell$ and $\sigma$ explicitly, we
give the rotational velocity and magnetic field strength at the outer boundary  to satisfy 
Equations~(\ref{eq:boundary_condition}) and (\ref{eq:boundary_condition2}).
Following \citet{Yamasaki_Yamada:2005}, we assume that a spherical shell is rotating
rigidly at a radius of $\mathrm{1000km}$ as
   \begin{eqnarray}
    \label{eq:rotation_law}
     \uph(r_{\mathrm{1000km}}, \theta) = 2\pi r_{1000\mathrm{km}} \sin \theta f_{1000\mathrm{km}}.
   \end{eqnarray}   
   Since the specific angular momentum $\ell$ is conserved along an individual streamline, 
   the rotational velocity just above the shock surface is given as
   \begin{eqnarray}
    \label{eq:rotation_law2}
    \uph(r, \theta) = 2\pi \frac{r_{1000\mathrm{km}}^2}{r} \sin \theta f_{1000\mathrm{km}},
   \end{eqnarray}
   if the streamlines are assumed to be radial. 
   \citet{Yamasaki_Yamada:2005} obtained steady solutions only for 
   $f_{1000\mathrm {km}}=0.03 \, \mathrm{s^{-1}}$ (slow rotation)
   and $f_{1000\mathrm{km}} = 0.1 \,\mathrm{s^{-1}}$
   (fast rotation). We consider a larger variety than \citet{Yamasaki_Yamada:2005} in this paper.

   As for the toroidal magnetic field, we assume the following profile at the same
   radius of $\mathrm{1000km}$ as
   \begin{eqnarray}
    \label{eq:toroidal_law}
     \Bph (r_{\mathrm{1000km}}, \theta) &= B_{1000\mathrm{km}} \sin \theta,
   \end{eqnarray}
       where $B_{1000\mathrm{km}}$ givens the strength of the
       toroidal magnetic field there. Since the $\sigma$ in Equation~(\ref{eq:ell_sigma})
       is a conserved quantity, the toroidal magnetic field just outside the
       shock surface is determined as
       \begin{eqnarray}
	\label{eq:toroidal_law2}
    \Bph (r, \theta) &= B_{1000\mathrm{km}} \sqrt{\dfrac{r_{1000\mathrm{km}}}{r}} \sin \theta,
       \end{eqnarray}
       if we again assume a radial infall upstream the shock wave (see Appendix~\ref{sec:derivation}).
       Note that these functional forms in Equations (\ref{eq:rotation_law}) and (\ref{eq:toroidal_law})
    satisfy the boundary conditions in Equations (\ref{eq:boundary_condition}) and
    (\ref{eq:boundary_condition2}).

 If the magnetic field has both poloidal and toroidal components,
 both of them should be fixed at the outer boundary. If we again assume 
 a radial free fall outside the shock, the poloidal magnetic field lines should be also radial. Then
 the continuity equation of the magnetic field gives 
 the poloidal magnetic field just outside the shock surface as
 \begin{eqnarray}
  \label{eq:poloidal}
   \Br(r,\theta) = \frac{r^2_{\mathrm{1000km}}}{r^2} B_0, \,\,\, \Bth(r,\theta) = 0,
 \end{eqnarray}
    where $B_0$ is a constant giving the strength of the poloidal magnetic field at the
    radius of $\mathrm{1000km}$.
    When the poloidal magnetic field is non-zero,
    the specific angular momentum and the toroidal magnetic field
    are no longer independent of each other (\citealt{Fujisawa_et_al:2013})
    and 

     Equations~(\ref{eq:rotation_law2}) and (\ref{eq:toroidal_law2}) are not valid.
     The values of $\uph(r_s, \theta)$ and $\Bph(r_s,\theta)$ just ahead of 
     the shock surface are then obtained
     by solving Equations~(\ref{eq:mag_field_ph_2}) and (\ref{eq:mhd_flow}) numerically 
     in Appendix~\ref{sec:derivation}.
  
\subsection{Numerical scheme and parameter settings} \label{ssec:parameter}

We summarize our new numerical scheme to obtain steady solutions briefly here.
More details are given in Appendix~\ref{sec:flowchart}.
Each steady solution is characterized by
five parameters, $\dot{M}$, $L_\nu$, $f_{1000\mathrm{km}}$, $B_{1000 \mathrm{km}}$ and $B_0$.
We discretize Equations~\eqref{eq:basics_hydro}~and~\eqref{eq:basics_mag} and 
obtain non-linear systems of equations. Setting the above five parameters and
assuming the values of $r_s$ and $\epsilon$ in Equation~(\ref{eq:rs}),
we solve these equations numerically from the
shock surface to down the neutrino sphere. If the density $\rho$ at the neutrino
sphere so obtained does not satisfy the requirement $\rho = 10^{11}~\mathrm{g~cm^{-3}}$,
we modify the values of $r_{\mathrm{s}}$ and $\epsilon$ and repeat the procedure.
$r_s$ and  $\epsilon$ are hence regarded as the eigenvalues in the boundary-value problem for this system.
In order to obtain the value of the critical neutrino luminosity,
we calculate a sequence of solutions for fixed
values of four parameters  ($\dot{M}$, $f_{1000\mathrm{km}}$, $B_{1000 \mathrm{km}}$ and $B_0$),
changing the value of $L_\nu$ until a steady state solution is no longer obtained. 
We develop a new numerical method dubbed the W4 method (\citealt{W4}), which is meant
to solve the non-linear systems of equations.
The details of the W4 method are described
in Appendix \ref{sec:W4} (see also our recent paper \citealt{W4}). 

Information on the rotation and  magnetic field deep inside the progenitor is scarce. 
In general, massive stars, such as B and O type stars, tend to have
rapid rotation (e.g.~\citealt{Hunter_et_al:2008}).
Approximately 10 per cent of them have surface rotational velocities
larger than 300 $\mathrm{km~s^{-1}}$ (\citealt{Ramirez-Agudelo_et_al:2013}).
On the other hand, recent surveys for massive stars indicate that 
some Galactic O and B type stars have magnetic fields of 100 -- 1000$~\mathrm{G}$ at
their surfaces (\citealp{Wade:2015}).
Their total magnetic fluxes roughly coincidence with those of a typical magnetar
whose magnetic field is about $10^{14-15}~\mathrm{G}$ on their surfaces. 
According to the stellar evolution models, the toroidal magnetic field is
likely to be larger by orders of magnitude than the poloidal
one due to the differential winding inside the massive
stars (\citealt{Heger_Woosley_Spruit:2005}). Some progenitors may hence
have a rapidly rotating and/or strongly magnetized core.

On the other hand, the stellar evolution models also indicate that
the transport of angular momentum during the quasi-static phase of
the progenitor reduces the angular momentum of the core,
particularly if the magnetic torque is taken into account (e.g.~\citealp{Heger_Woosley_Spruit:2005};
\citealp{Potter_et_al:2012}). The rotation might not be so rapid after all then. 
If this is really the case, rotation does not play an important role in the dynamics of core collapse.
One should bear in mind, however, that almost all stellar evolution
 calculations are based on spherically
averaged one-dimensional models and have uncertainties in the mechanism and formulation
of angular momentum transport and magnetic field. 

The aim of this paper is to systematically study the effects of rotation and magnetic field on the
critical neutrino luminosity in CCSNe using the
new numerical scheme. We consider rapid rotation and/or strong toroidal
magnetic field, setting the parameters as $f_{1000\mathrm{km}} \sim 0 - 0.45~\mathrm{s^{-1}}$,
$B_{1000\mathrm{km}} \sim 0 - 3.0 \times 10^{12}~\mathrm{G}$ and $B_0 \sim 0 - 10^{11}~\mathrm{G}$.
 These values roughly correspond to
$f\sim 0 - 5~\mathrm{ms^{-1}}$ and $B\sim 0 - 5 \times10^{14}~\mathrm{G}$
on the PNS surface, which are the typical rotation frequency of a milisecond pulsar
and the canonical strength of the surface magnetic field of a magnetar respectively.
The PNS neutrino temperature and the mass 
are fixed to $T_\nu = 4.5~{\mathrm{MeV}}$ and $M = 1.3~\mathrm{M_\odot}$, respectively in this paper.

\section{Numerical results}\label{sec:results}

\subsection{Streamlines of steady accretion flow}

\begin{figure*}[t]
 \begin{tabular}{cc}
  \includegraphics[width=8cm,clip]{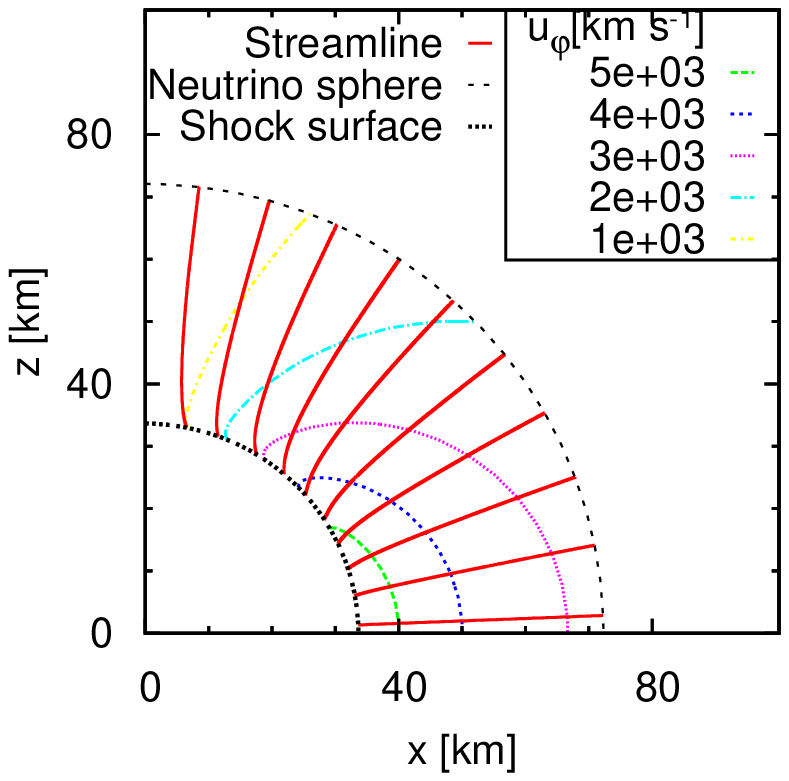}&
  \includegraphics[width=8cm,clip]{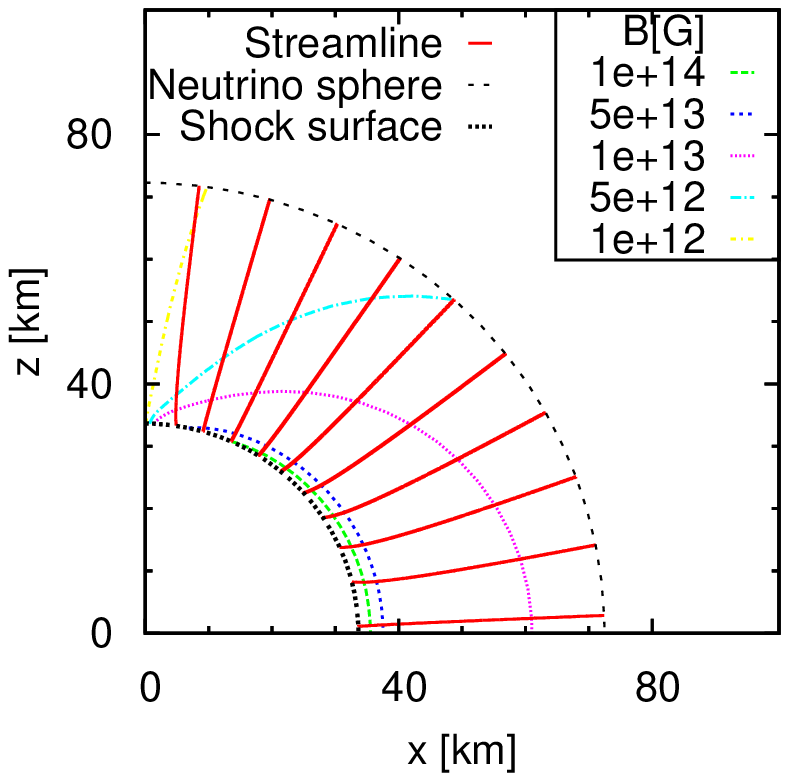}
 \end{tabular}
 \caption{
   Streamlines in the meridian section  (solid red lines)
   for the models rotation (left) and toroidal magnetic field (right).
   The mass accretion rate and neutrino luminosity are set to 
   $\dot{M}=0.5~\mathrm{M_{\odot}~s^{-1}}$ and $L_{\nu}=26\times 10^{51}~\mathrm{erg~s^{-1}}$,
   respectively. The innermost black-dotted curve is the
   neutrino sphere and the outermost black-dashed curve indicates the shock surface
   in each panel. In the left panel the dash-dotted lines denote the profile of the
   rotational velocity $\uph$; the rotational frequency
   is $f_{1000\mathrm{km}} = 0.2~\mathrm{s^{-1}}$ and
   the value of $\epsilon$ for shock deformation
   is $\epsilon = 2.0\times 10^{-3}$. In the right panel the 
   dash-dotted lines show the profile of the toroidal magnetic field  $\Bph$;
   the strength of the toroidal magnetic field at the outer boundary
   is $B_{1000\mathrm{km}}= 2\times 10^{12}~\mathrm{G}$ and 
   $\epsilon = 1.8\times 10^{-3}$.}

 \label{fig:streamline}
\end{figure*}
 \begin{figure*}
  \begin{tabular}{cc}
   \includegraphics[width=8cm,clip]{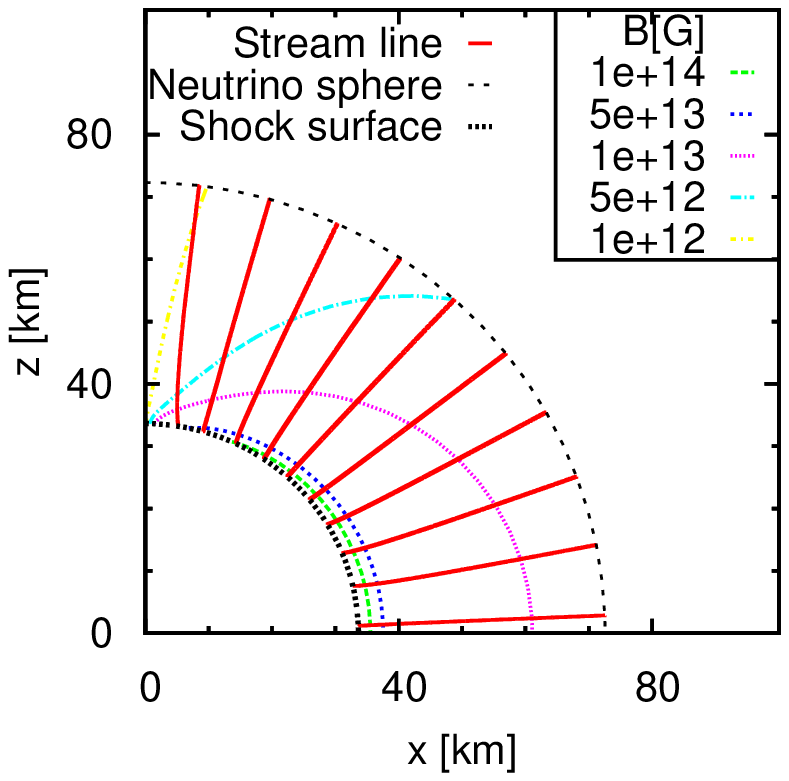}
   \includegraphics[width=8cm,clip]{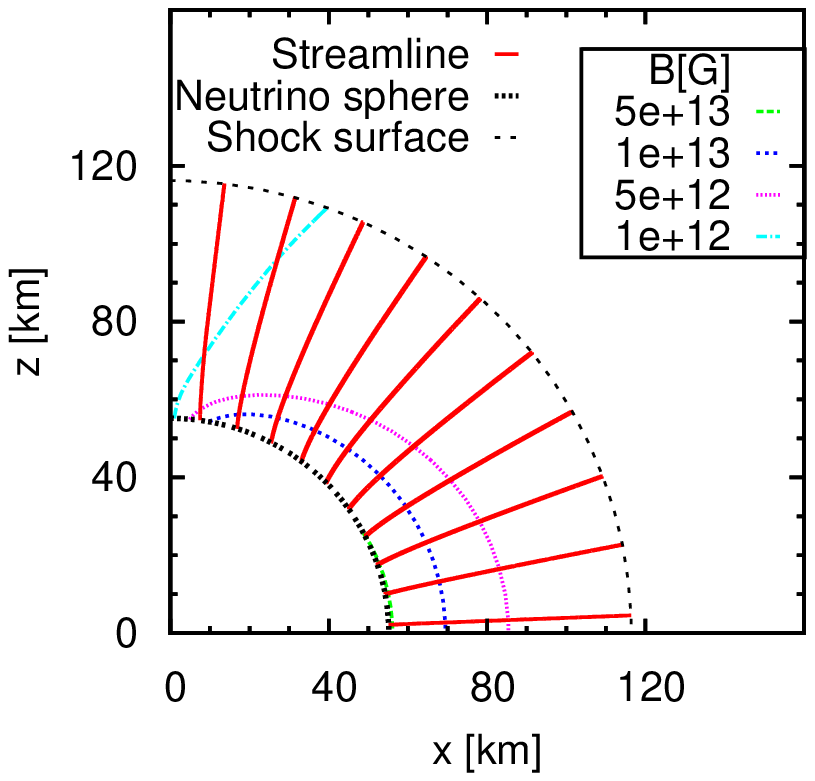}&
  \end{tabular}
  \caption{Same as in Figure \ref{fig:streamline} but for a model with
  different rotation and toroidal magnetic field (left) and for a model 
  with poloidal magnetic field in addition (right).
  In the left panel the mass accretion rate and neutrino luminosity are set to 
  $\dot{M}=0.5~\mathrm{M_{\odot}~s^{-1}}$ and $L_{\nu}=26\times 10^{51}~\mathrm{erg~s^{-1}}$,
  respectively. The rotational frequency is $f_{1000\mathrm{km}}=0.1~\mathrm{s^{-1}}$ and
  the strength of the toroidal magnetic field at the outer boundary is
  $B_{1000\mathrm{km}}= 2\times 10^{12}~\mathrm{G}$. 
  The value of $\epsilon$ for shock deformation is found to be $\epsilon = 2.3\times 10^{-3}$.
  In the right panel the mass accretion rate and neutrino luminosity are set to 
  $\dot{M}=2.0~\mathrm{M_\odot~s^{-1}}$ and $L_\nu=70 \times 10^{51}~\mathrm{erg~s^{-1}}$. 
  The rotational frequency is $f_{1000\mathrm{km}} = 0.2~\mathrm{s^{-1}}$ and 
  the strengths of the toroidal and poloidal
  magnetic fields are given as $B_{1000\mathrm{km}} =10^{12}~\mathrm{G}$ and 
   $B_0=10^{11}~\mathrm{G}$ at the outer boundary. We find again that 
  the value of $\epsilon$ for shock deformation is  $\epsilon = 2.3 \times 10^{-3}$.
  The poloidal magnetic field lines are coincident with the streamlines
  because of the ideal MHD condition.  
  }
  \label{fig:streamline_rotmag}
 \end{figure*}
\begin{figure*}
  \begin{tabular}{cc}
   \includegraphics[width=8.4cm,clip]{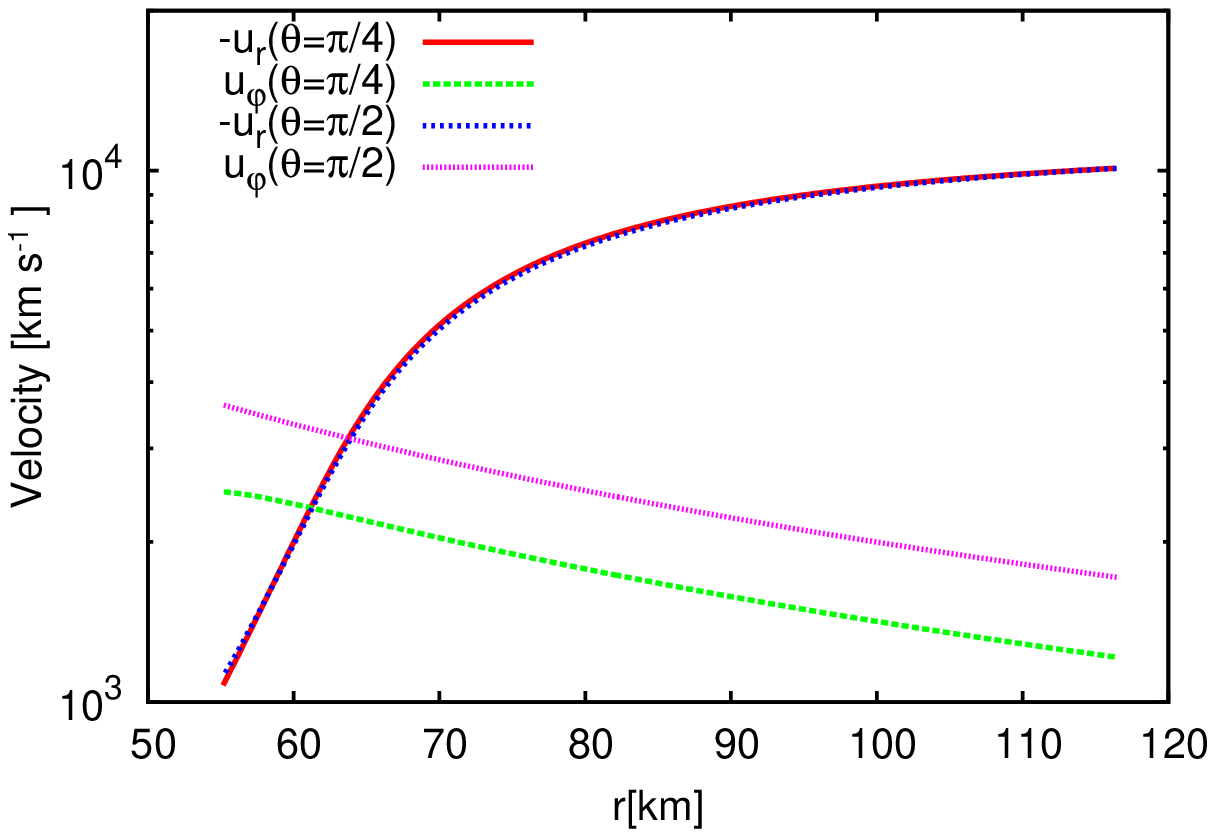}
   \includegraphics[width=8.4cm,clip]{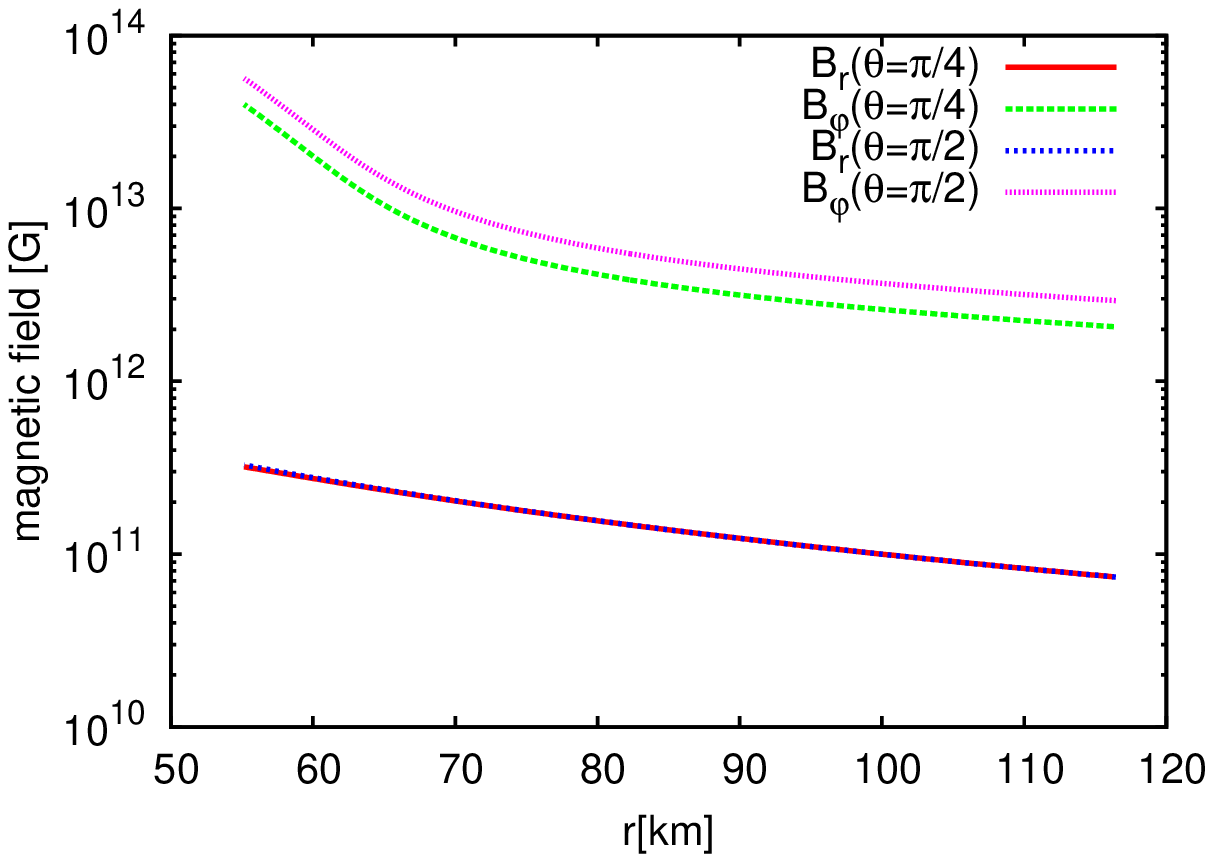}   
  \end{tabular}
 \caption{Radial profiles at $\theta=\pi/4$ and $\pi/2$ 
 of the radial and azimuthal components of velocity (left) and magnetic field (right)
 of the model shown in the
 right panel of Figure~\ref{fig:streamline_rotmag}.
 Both the poloidal and toroidal magnetic fields are taken into account. 
 }
 \label{fig:poloidal}
\end{figure*}

Figure \ref{fig:streamline} displays streamlines in a meridian plane for a model
either with rotation (left panel) or with toroidal magnetic field (right panel).
In these models the neutrino luminosity and accretion rate are set
as $L_\nu = 26 \times 10^{51}~\mathrm{erg}~\mathrm{s}^{-1}$
and $\dot{M}=0.5~\mathrm{M}_{\odot}\,\mathrm{s^{-1}}$.
The innermost black-dotted curve indicates the neutrino sphere
whereas the outermost black-dashed one corresponds to the stalled shock surface.
We also draw the contour lines for the azimuthal components of velocity
and magnetic field in the left and right panels, respectively.
The values of $r_s$ and $\epsilon$ (see Equation~{\ref{eq:rs}) are
$r_s \sim 70$~km and $\epsilon \sim 10^{-3}$ respectively,
in both models. Interestingly, although the shapes of the
shock surfaces are almost the same, being oblate,
the flow pattern are different from each other.
In fact in the left panel 
of Figure~\ref{fig:streamline},
the flow is pushed toward the equatorial plane
as it approaches the PNS. This is due to
centrifugal force and was previously seen in Figure 6 in \citet{Yamasaki_Yamada:2005}.
In the right panel, on the other hand, the flow is bent toward the
symmetry axis. This is due to Lorentz force
exerted by the toroidal magnetic field,
which indeed behaves as a negative centrifugal force (e.g.~\citealt{Fujisawa:2015a}).
Noted that the flow patterns of both models are similar near the rotation axis
 because of the boundary condition for $\uth$ at the axis (Equation~\ref{eq:boundary_condition}).

The left panel in Figure \ref{fig:streamline_rotmag}
shows the result for a model with  both rotation and toroidal magnetic field.
The mass accretion rate, neutrino luminosity and
strength of toroidal magnetic field are set to 
 the same as values in Figure~\ref{fig:streamline},
 but the rotation frequency is somewhat smaller.
 We find $\epsilon = 2.3\times10^{-3}$, slightly larger than in the previous cases
 in Figure~\ref{fig:streamline}. The centrifugal force and the Lorentz force
 almost cancel each other in this solution.
 The streamlines are nearly 
 radial in this model except for the inner region where the Lorentz force is dominant over
 the centrifugal force and the streamlines look similar to
 those for purely toroidal magnetic field.
 This is understood from the conserved quantities $\ell$ and $\sigma$
 along the streamline.
 As a matter of fact, $\ell$ does not depend on the density
 while $\sigma$ does as is clear in Equation (\ref{eq:ell_sigma}).
 Since the density is low near the shock surface $\rho \sim 10^{9}~\mathrm{g~cm^{-3}}$
 and gets higher toward the neutrino sphere, where 
 $\rho \sim 10^{11}~\mathrm{g~cm^{-3}}$.
 The Lorentz force tends to be dominant in the inner region and vice versa. 
 
 The right panel in Figure~\ref{fig:streamline_rotmag} shows streamlines for a model
 that incorporates poloidal magnetic fields in addition to toroidal ones. 
The neutrino luminosity and accretion rate are set as $L_\nu = 70\times 10^{51}~\mathrm{erg~s^{-1}}$
and $\dot{M} = 2.0~\mathrm{M_\odot}~\mathrm{s}^{-1}$, respectively. 
 The rotational frequency and the strength of magnetic fields are
 $f_{1000\mathrm{km}} = 0.2~\mathrm{s^{-1}}$,
$B_{1000\mathrm{km}} = 10^{12}~\mathrm{G}$ and $B_0 = 10^{11}~\mathrm{G}$, respectively.
Since the value of $B_0$ is smaller by an order than that of $B_{\mathrm{1000km}}$,
the poloidal magnetic field in this model is much weaker than the toroidal magnetic field.
We find that the value of $\epsilon$ is $\epsilon = 2.3 \times 10^{-3}$
again. Note that the poloidal magnetic field lines
are aligned with the streamlines in ideal MHD.
They are slightly curved near the neutrino sphere
similarly to the left panel in Figure~\ref{fig:streamline}
because of Lorentz force mainly exerted by the toroidal magnetic field. 

Figure~\ref{fig:poloidal} displays the radial and the azimuthal components of
velocity (left) and magnetic field (right) at $\theta=\pi/4$ and $\theta=\pi/2$.
The radial components of the flow velocities $\urr$
and of the poloidal magnetic field $\Br$ are almost identical at these zenith angles.
This is mainly because of our assumption that the flow and the poloidal magnetic field
outside the shock surface are radial and 
independent of $\theta$. If we had assumed a $\theta$-dependent
functional form in Equation \eqref{eq:poloidal}, for example,
we would have found $\theta$-dependent radial profiles of $\Br$.
In contrast, both $\uph$ and $\Bph$
depend on $\theta$, being larger on equator ($\theta = \pi/2$) than at $\theta = \pi / 4$.
This $\theta$ dependence again ($\propto \sin \theta$), however, simply reflects
of the functional forms for $\uph$
and $\Bph$ in Equations~(\ref{eq:rotation_law}) and (\ref{eq:toroidal_law}).

The rotational velocity and toroidal magnetic field both increase inwards 
from the shock surface to the neutrino sphere.
It is apparent, however, that the toroidal magnetic field rises more steeply than the
rotational velocity.
 This trend is almost independent of the functional forms for $\uph$ and $\Bph$.
 In fact, it is dictated by the conservation of $\ell$ and $\sigma$,
 the latter of which depends on the density profile as is explicit in Equation (\ref{eq:app_ell_sigma2})
 in Appendix~\ref{sec:derivation}.
 One may hence roughly say that
 the $\theta$ dependence of the flow velocities and magnetic fields is mainly 
 determined by their dependence just ahead of the shock wave,
 while their radial profiles are largely constrained by the conservation laws 
(Equation \ref{eq:app_ell_sigma2}), being almost independent
of the outer boundary conditions.

\subsection{Effects of rotation and magnetic field on the critical luminosity}

\begin{figure*}
 \begin{tabular}{cc}
  \includegraphics[width=9cm]{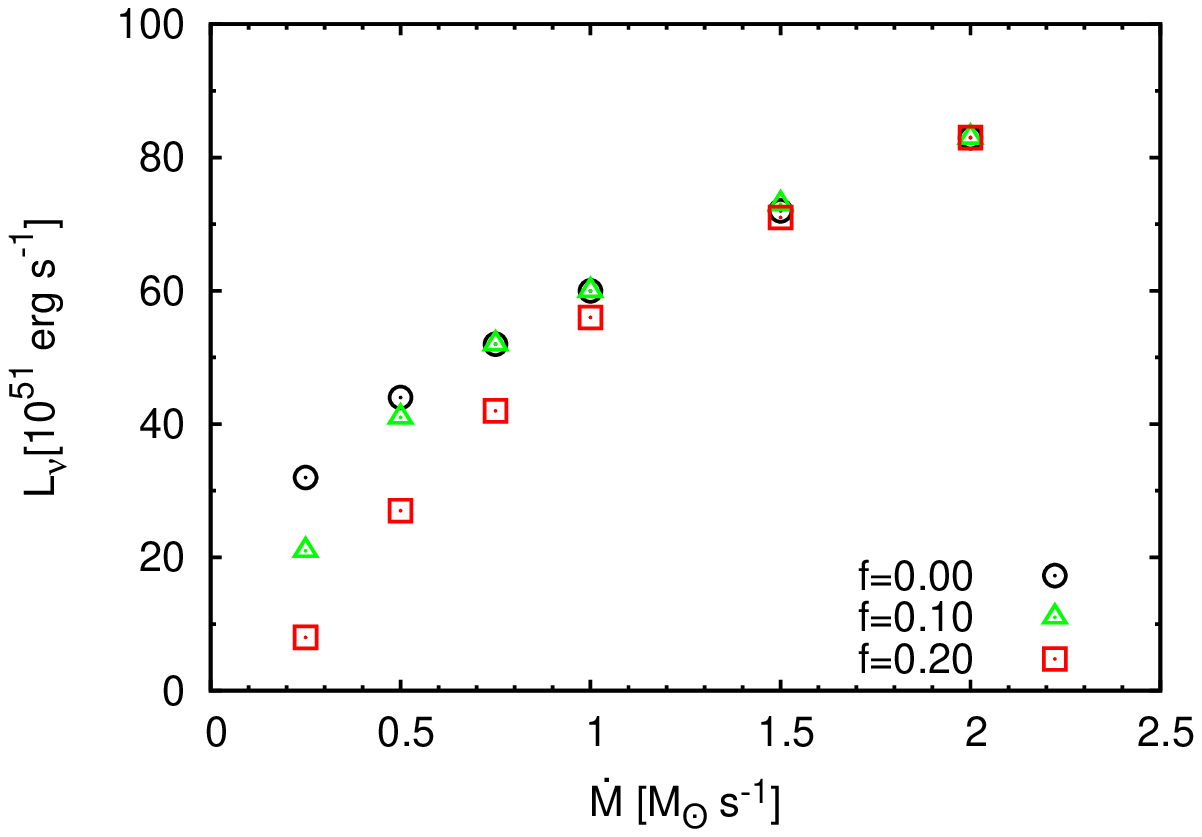}
  \includegraphics[width=9cm]{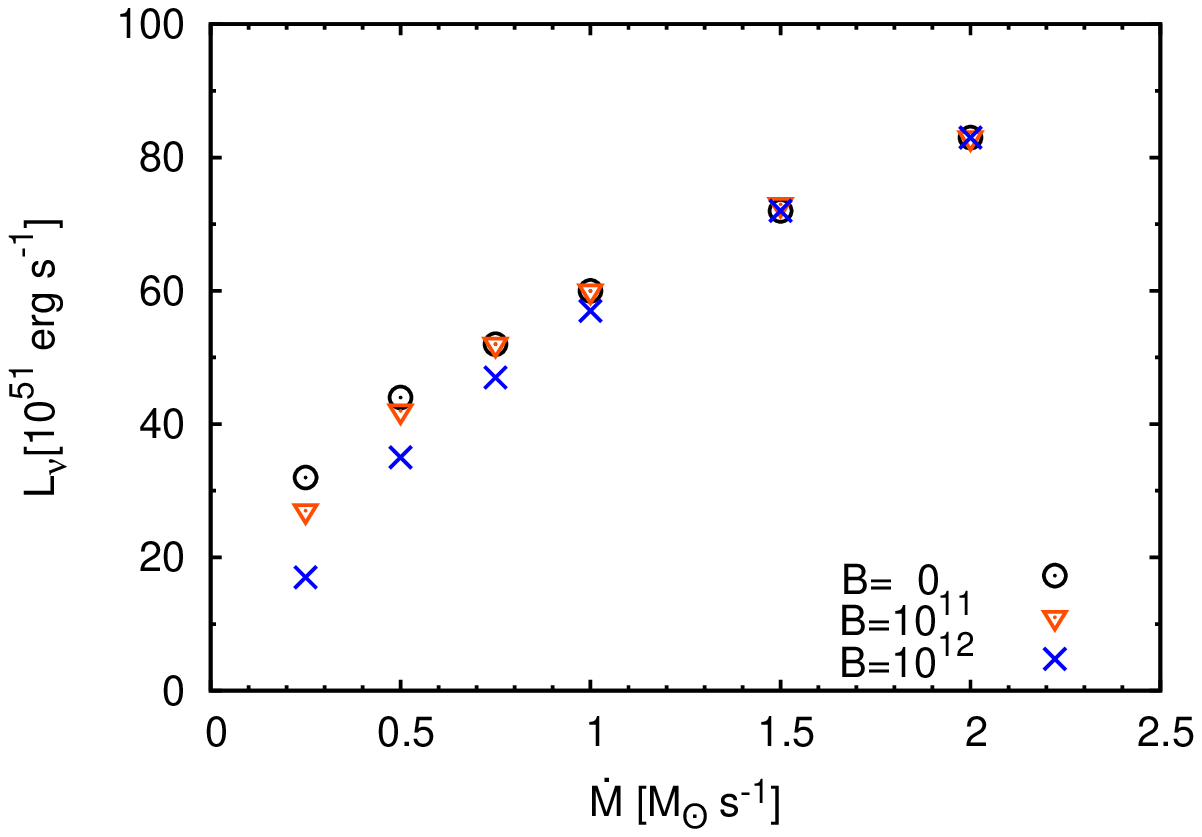}
 \end{tabular}
 \caption{
 The critical luminosity as a function of the mass accretion rate
 for the models with rotation alone (left) and with purely toroidal magnetic field alone (right).
 The black circles represent the spherical models
 without rotation and magnetic field. 
 In the left panel green triangle and red squares correspond to 
 rotation rates
 $f_{1000\mathrm{km}} = 0.1~\mathrm{s^{-1}}$
 and $f_{1000\mathrm{km}} = 0.2~\mathrm{s^{-1}}$, respectively. 
 In the right panel field strength of $B_{1000\mathrm{km}} = 10^{11 }~\mathrm{G}$
 and $B_{1000\mathrm{km}} = 10^{12}~\mathrm{G}$, respectively.
 }
 \label{fig:critical_luminosity} 
\end{figure*}

\begin{figure*}
 \begin{tabular}{cc}
  \includegraphics[width=9cm]{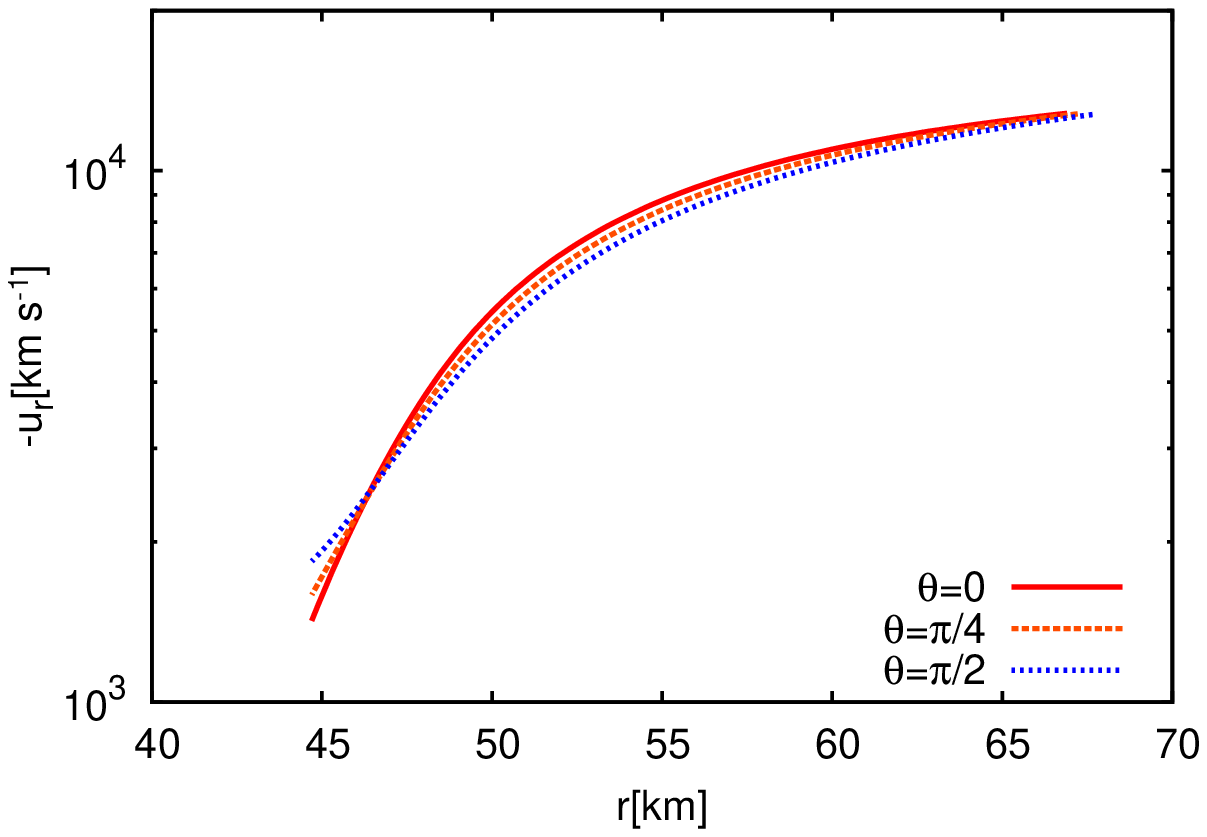}
  \includegraphics[width=9cm]{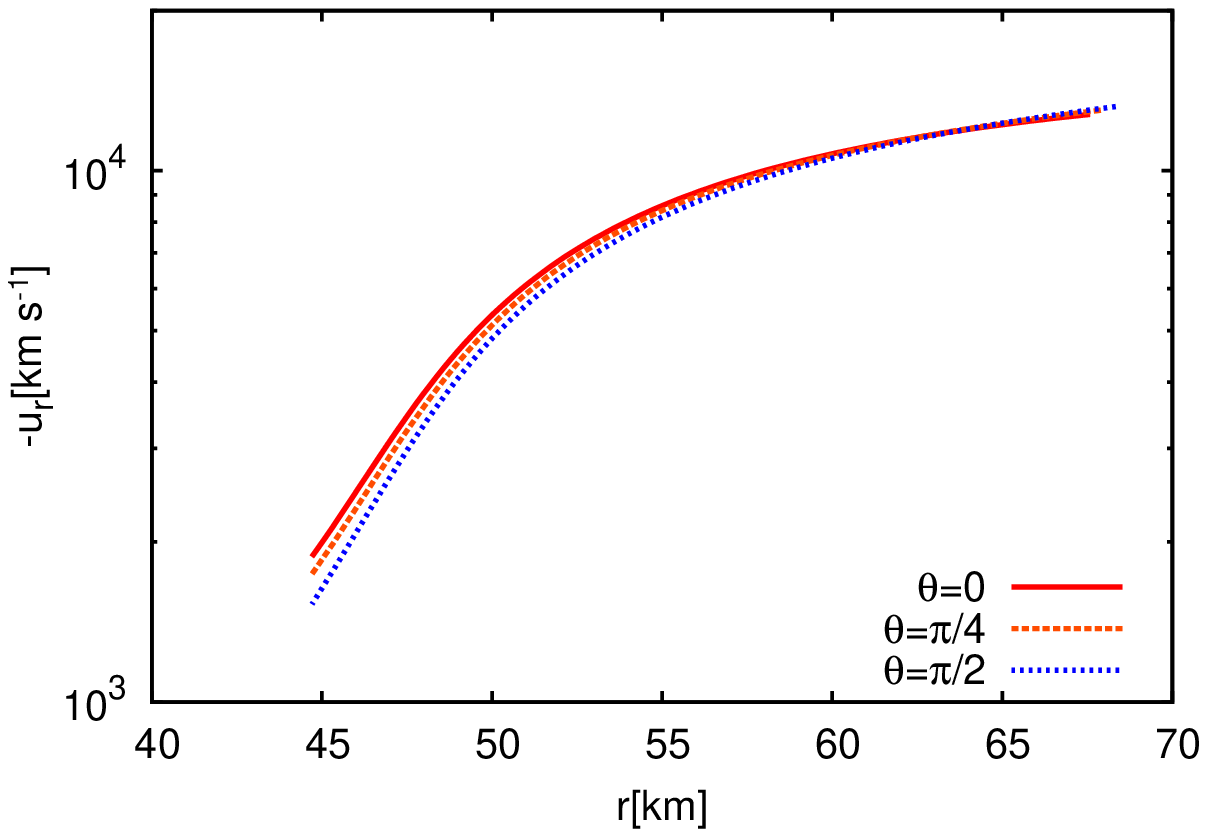}
 \end{tabular}
 \caption{
 Radial profiles at $\theta = 0$ (red solid lines), $\pi/4$ (red dotted lines)
 and $\pi/2$ (blue dotted lines) of the radial velocity for ones of the typical solutions 
 at the critical neutrino luminosities for the models with rotation alone
 (left) and for the models with purely toroidal magnetic field alone (right panel).
 In the both panels
 the mass accretion rate and neutrino luminosity are set to 
 $\dot{M}=2.0~\mathrm{M_{\odot}~s^{-1}}$ and $L_{\nu}=46\times 10^{51}~\mathrm{erg~s^{-1}}$,
 respectively. The rotational frequency is $f_{\mathrm{1000km}} = 0.5~\mathrm{s^{-1}}$ (left) and
 the strength of the toroidal magnetic field is
 given as $B_{1000\mathrm{km}} = 9 \times 10^{12}~\mathrm{G}$ (right).
 }
 \label{fig:radial_velocity} 
\end{figure*}

\begin{figure*}
 \includegraphics[width=7cm,clip]{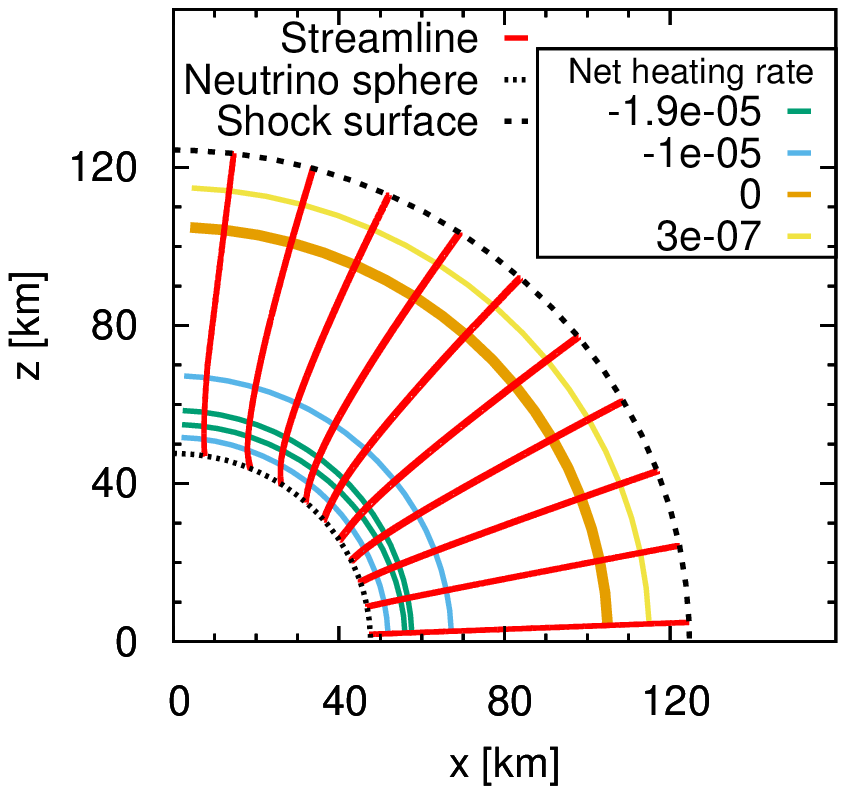}
 \includegraphics[width=10cm,clip]{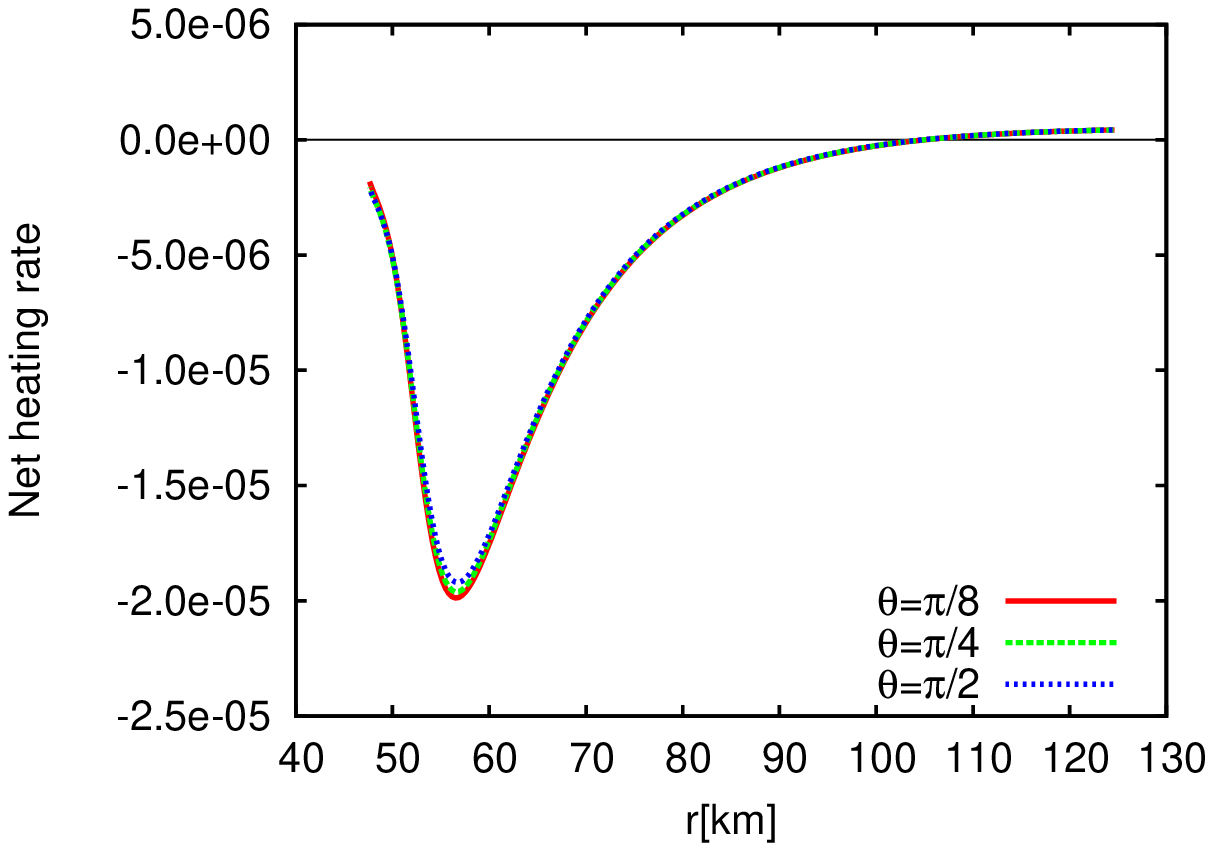}
 \caption{Streamlines (left) and net heating rates along some radial ray (right)
 in the meridian section for one of the purely rotational models at
 its critical luminosity. 
 The mass accretion rate and neutrino luminosity are set to $\dot{M}=1.0~\mathrm{M_{\odot}~s^{-1}}$, 
 $L_{\nu}=52\times 10^{51}~\mathrm{erg~s^{-1}}$, respectively.
 The rotational parameter is $f=0.2~\mathrm{s^{-1}}$ and 
 the neutrino-heating rate is normalized by $c^3$.
 In the left panel the thick curve indicates the boundary between the heating region and the cooling region.
 In the right panel the three lines correspond to different zenith angles:
  $\theta = \pi/8$ (red solid line), $\theta = \pi/4$ (green dashed line)
 and $\theta = \pi/2$ (blue dotted line).}
 \label{fig:gain}
\end{figure*}
\begin{figure*}
 \includegraphics[width=7cm,clip]{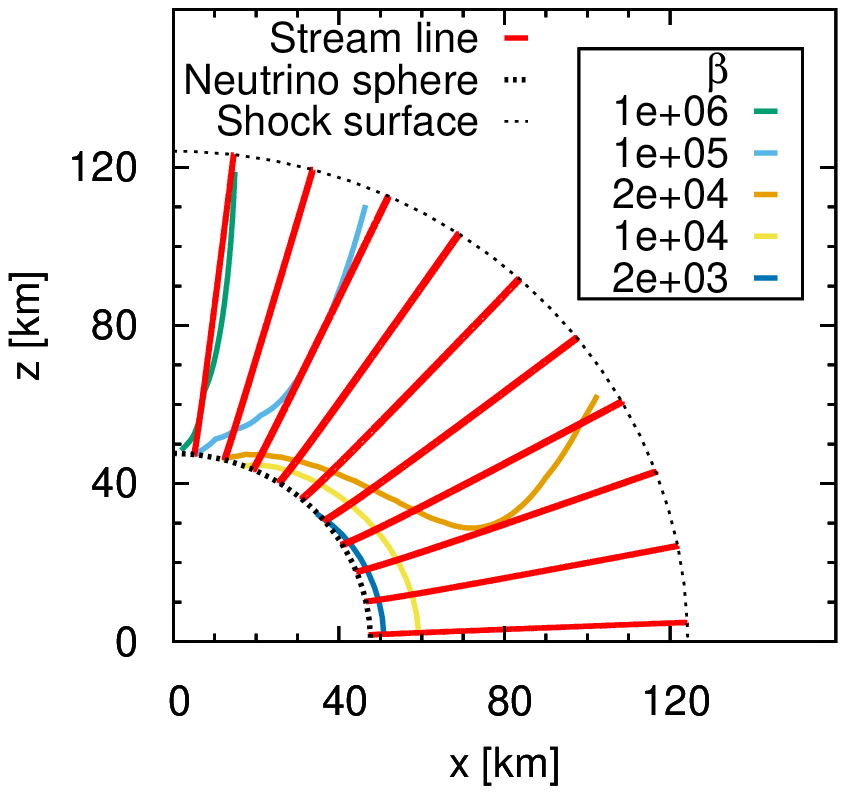}
 \includegraphics[width=10cm,clip]{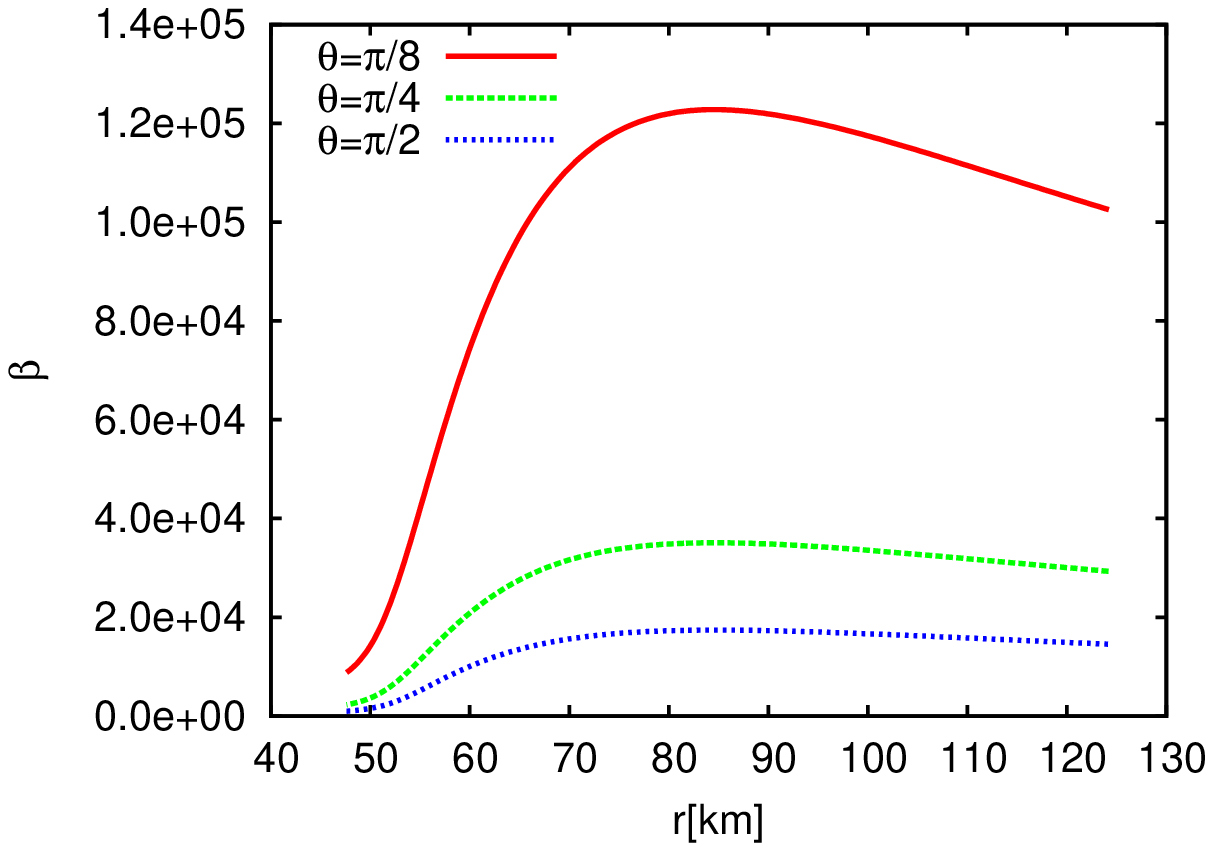}
 \caption{
 Streamlines (left) and radial profiles of plasma $\beta$ along some radial rays (right)
 for one of the purely toroidally-magnetized models close to its critical luminosity. 
 The mass accretion rate and neutrino luminosity are set to $\dot{M} = 1.0~\mathrm{M_{\odot}~s^{-1}}$,
 $L_{\nu}=52\times 10^{51}~\mathrm{erg~s^{-1}}$  and $B_{1000\mathrm{km}}=10^{12}~\mathrm{G}$, respectively.
 In the right panel the three lines correspond to different zenith angles:
 $\theta = \pi/8$ (red solid line), $\theta = \pi/4$ (green dashed line)
 and $\theta = \pi/8$ (blue dotted line).
 }
 \label{fig:plasma_beta}
\end{figure*}

\begin{figure*}
 \begin{tabular}{cc}
 \includegraphics[width=9cm]{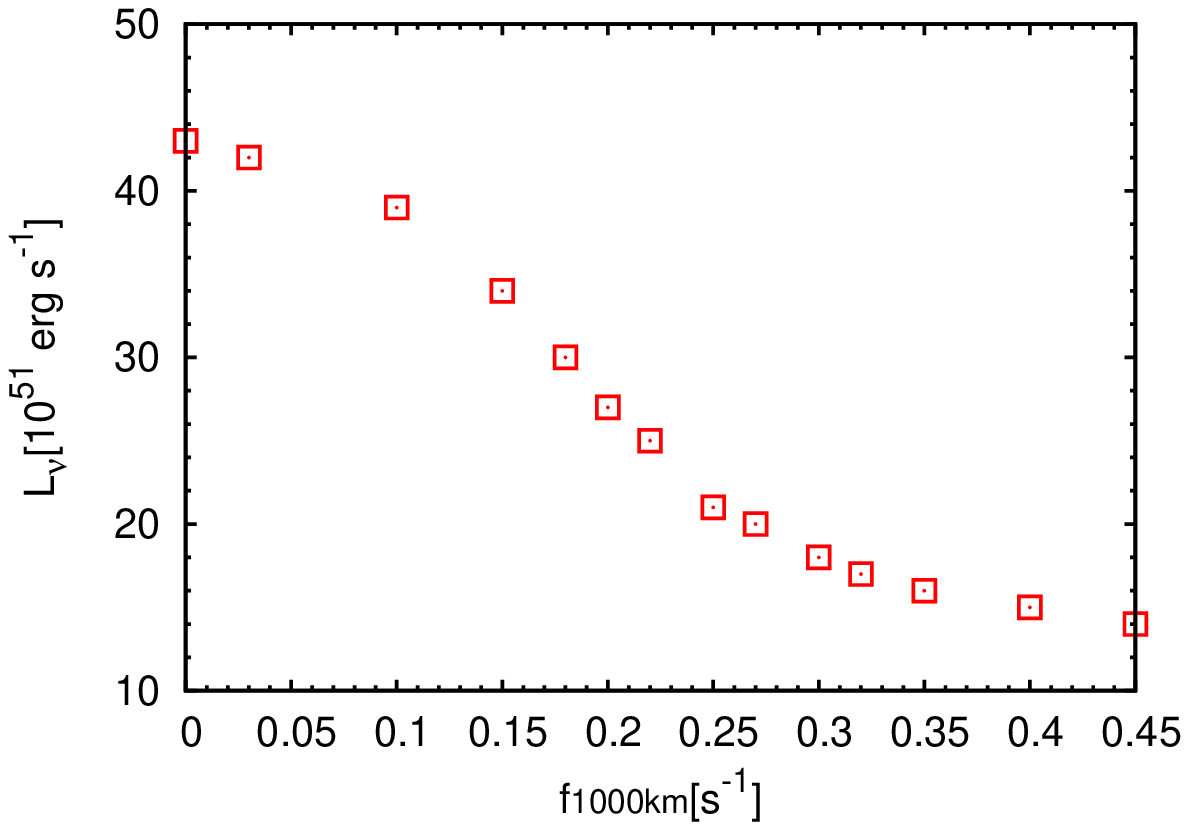}&
 \includegraphics[width=9cm]{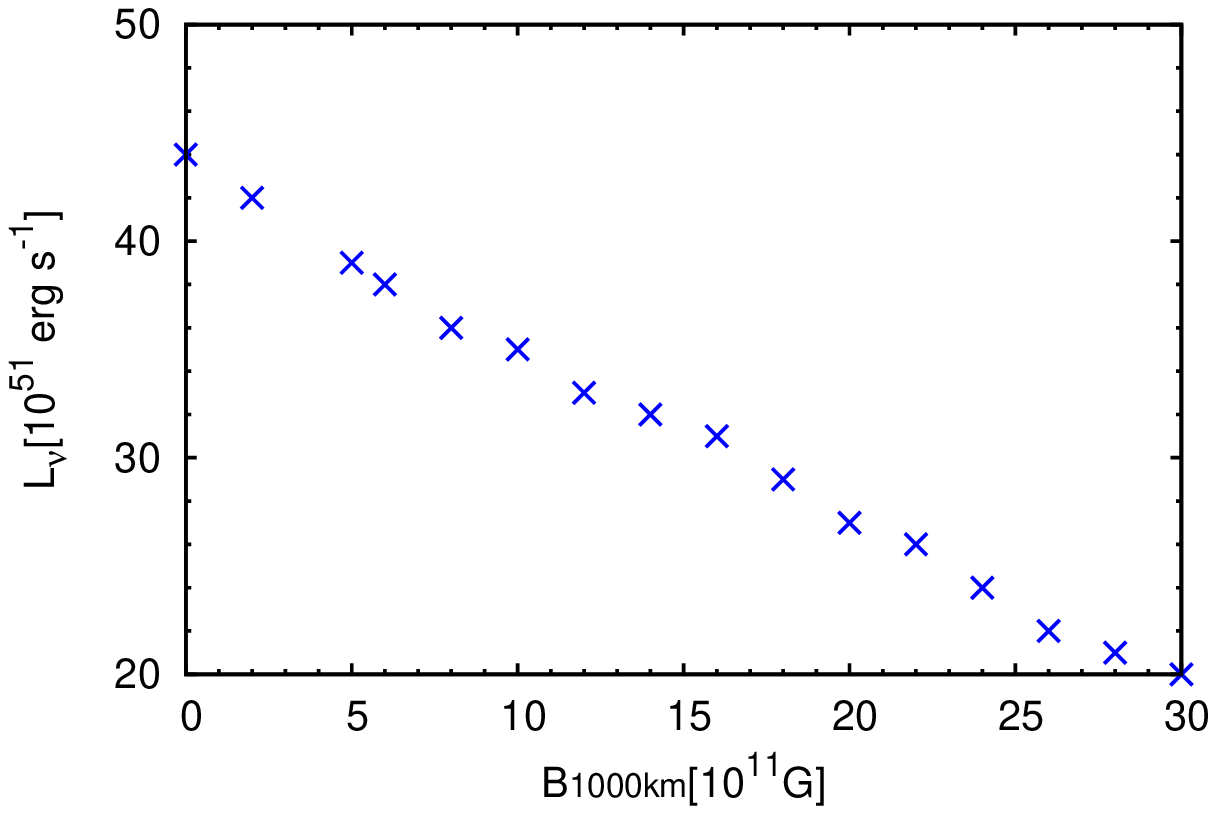} 
 \end{tabular}
 \caption{The critical luminosity as a function of the rotational rate $f_{1000\mathrm{km}}$ (left)
 and the strength of the toroidal magnetic field (right).
 The mass accretion rate is fixed to $\dot{M} = 0.5 \mathrm{M_\odot~s^{-1}}$.
 }
 \label{fig:Lnu_vs_f}
\end{figure*}

Now we look into the changes in the 
critical luminosity that rotation and/or magnetic field produce.
In this subsection we neglect poloidal magnetic fields 
($B_0 = 0$) because the toroidal magnetic field is supposed to be dominant.
Figure \ref{fig:critical_luminosity} displays the critical luminosities
either for models with rotation alone (left panel) or for models with
toroidal magnetic field alone with no rotation (right panel),
which are plotted as a function of the mass accretion rate.
As a reference, we also plot the critical luminosity of the spherical symmetric models
 without rotation and toroidal magnetic field as black circles.

It is evident among other things that 
the critical luminosity for all these models
with either rotation or toroidal magnetic field 
are lower than those for the spherical models.
Non-spherical forces, i.e., centrifugal forces and hoop
stresses, tend to reduce the critical luminosity
although the reduction rate depends on the mass accretion rate. 
As a matter of fact, for a given rotation velocity or magnetic field strength
at the outer boundary,
the critical luminosity
gets lowered more strongly as the mass accretion rate becomes smaller:
the reduction rate is as high as 
about $50\%$ -- $70\%$ for the rapid rotation with $f_{\mathrm{1000 km}} = 0.2$
and $20\%$ -- $50\%$ for the strong toroidal magnetic field with $B_{\mathrm {1000 km}} = 10^{12}~\mathrm{G}$ 
at the mass accretion rates lower than $1.0~\mathrm{M_\odot}$.

Figure \ref{fig:radial_velocity} displays the radial
profiles of the radial velocity $\urr$ along three radial rays with $\theta = 0$ (red solid line),
$\pi / 4$ (orange dotted line) and $\pi/2$ (blue dashed line) for representative models on the
critical curve either with a rotation alone (left panel) or with a toroidal magnetic field alone (right panel).
The three profiles are almost the same at large radii in both panels whereas 
they are different from each other in the inner region.
In the left panel for pure rotation, 
the radial infall velocity decreases more rapidly near the neutrino sphere along the rotation
axis ($\theta = 0$) than on the equator ($\theta = \pi/2$).
This tendency was already observed in the previous work (\citealt{Yamasaki_Yamada:2005}). 
The radial profile of $\urr$ on the equator is convex at large radii but turns concave 
near the neutrino sphere whereas it remains convex along the polar axis as in the spherically
symmetric case. Considering the fact that in the spherical symmetry 
the profile of the radial velocity is still convex near the inner boundary
at the critical luminosity (see Figure 1 in \citealt{Yamasaki_Yamada:2005}),
one may say that only the polar flow reaches 
the critical state locally and shock revival will commence there. 

Interestingly, the situation is other way around  
in the right panel for the purely 
toroidal magnetic field.
This is because such fields exert hoop stress, which behaves like a
negative centrifugal force. As a result, 
the radial inflow velocity is smaller on the equator than on the axis 
near the neutrino sphere.
According to the same argument, this may imply that only the equatorial flow
becomes critical state locally and shock revival will be initiated on the equator.
It is intriguing that the existence/non-existence of steady solutions that satisfy a
certain boundary condition is itself a global issue while 
the critical state is realized locally as a result.
Although some recent studies strongly deduce the presence of global conditions under critical state
 in spherical symmetry (e.g. \citealt{Murphy_Dolence:2017}),
one may hence claim that the critical neutrino luminosity is determined both globally and locally. 

Figure \ref{fig:gain} displays the streamlines (left panel) and
the radial profiles of the net heating rates along some radial rays (right panel) in the meridian
section for one of the purely rotational models at its critical luminosity. 
The thick orange curve in the left panel indicates the boundary between the cooling and heating regions,
or the 2D analog of the gain radius. 
One finds in the right panel that the
cooling region near the pole (see the green line) is wider and deeper than that on the equator
although the difference is not very large. Noted that for simplicity 
we assume in this paper that the neutrino sphere is
spherical (see also Equation~\ref{eq:qdot}}), which is certainly not true and 
tends to reduce the difference.
We emphasize, however, that this slightly
deformed cooling region results in the reduction of the critical neutrino luminosity
as observed earlier. 

 Figure~\ref{fig:plasma_beta} shows distribution of the plasma $\beta$
 defined as
 \begin{eqnarray}
 \beta = \frac{8 \pi p}{|\Bph|^2}.
 \end{eqnarray}
 in the meridian section for one of the non-rotating but purely
 toroidally-magnetized models close to its critical luminosity.
 The value of $\beta$ is very high, $> 10^3$, almost everywhere in the computational
 domain, meaning that the magnetic pressure is much smaller than the matter pressure
 except near the equatorial area on the neutrino sphere where the value of $\beta$ is
 much smaller. This locally strong magnetic pressure (or locally low $\beta$)
 modifies the accretion flow there and gives rise to the critical state 
 locally, leading to the reduction of the critical luminosity, as shown in the right
 panel of Figure~\ref{fig:radial_velocity}.

 Finally we show in Figure~\ref{fig:Lnu_vs_f}
 the critical luminosity as a function of the rotational rate 
 $f_{\mathrm{1000km}}$ (left) and the strength of the toroidal magnetic field
 $B_{\mathrm{1000km}}$ (right). 
 We fix the mass accretion rate as $\dot{M} = 0.5 \mathrm{M_\odot}\, \mathrm{s^{-1}}$.
 The critical luminosities of solutions with $f_{1000\mathrm{km}} = 0.25~\mathrm{s^{-1}}$
 or $B_{\mathrm{1000km}} = 3.0 \times 10^{12}~\mathrm{G}$
 are less than $20 \times 10^{51} \mathrm{erg~s^{-1}}$.
 The critical luminosity is reduced by $\sim 70\%$ for the most rapid rotation
 and by $\sim 50\%$ for the strongest toroidal magnetic field, respectively.
 It is interesting that we cannot find a steady state solution
 for the rotation parameter $f_{1000 \mathrm{km}} > 0.45~\mathrm{s^{-1}}$
 or the toroidal magnetic field strength $B_{1000\mathrm{km}} > 3.0\times10^{12}~\mathrm{G}$
 because the shock surface comes too close to 
 the neutrino sphere.
 These numerical results may imply that
 the existence of the critical specific angular momentum and the critical strength of the
 toroidal magnetic field, above which there actually exist no steady solutions.

\section{Discussion and summary}\label{sec:discussion}

We have derived numerically steady, non-spherical accretion flows through the
 standing shock wave onto the PNS solutions of stalled shock wave in the CCSNe core
 and effects of rotation and magnetic field on the shock revival in this paper.
In order to obtain these steady solutions, 
we have developed a new numerical scheme
to solve a system of nonlinear equations numerically, which we named the 
W4 method for non-linear equations,
and we have indeed succeeded in
generating various accretion flows with both rotation and magnetic field incorporated self-consistently
in axisymmetric 2D.
It should be noted that our new method can handle both poloidal and toroidal magnetic fields
simultaneously.
 
Our main findings are summarized as follows.
\begin{enumerate}
 \item The shock surface and the flow pattern become non-spherical
       by rotation and/or magnetic field in general. The shock surface
       is always oblate ($\epsilon > 0$)
       whereas the stream lines are bent either toward the 
       equatorial plane by rotation and poloidal magnetic field
       or toward the pole plane by toroidal magnetic field
       (Figures~\ref{fig:streamline} and~\ref{fig:streamline_rotmag}).

 \item The toroidal magnetic field is dominant in the inner region (Figure~\ref{fig:plasma_beta})
       while the rotation is dominant
       in the outer region (Figures~\ref{fig:streamline_rotmag} and \ref{fig:poloidal})
       because of the conserved quantities $\sigma$ and $\ell$ in Equations~\eqref{eq:ell_sigma}
       and \eqref{eq:app_ell_sigma2}.

 \item The critical luminosity is lowered by rotation and/or magnetic field in general
       although the degree of reduction depends on the mass accretion rate (Figure~\ref{fig:critical_luminosity}).

 \item In the presence of rotation and/or magnetic field the 
       critical state is realized locally: either on the symmetry axis for rotation
       or on the equatorial plane for toroidal magnetic field (Figure~\ref{fig:radial_velocity})
       despite the solution itself is determined globally
       according to the boundary conditions.

 \item The gain region in the accretion flow is also deformed.
       The cooling region is widened near the pole by the rotation (Figure~\ref{fig:gain})
       whereas it is affected mostly near the equator by the toroidal magnetic field
       (Figure~\ref{fig:plasma_beta}).
       Although the deviation from the spherical symmetry is small, it 
       results in the substantial reduction of the critical neutrino luminosity.

 \item It is suggested the existence of the non-vanishing 
       critical specific angular momentum,
       above which no steady solution exists irrespective of the neutrino luminosity
       (left panel in Figure~\ref{fig:Lnu_vs_f}).
       Our results are roughly consistent with
       the dynamical simulations in  \citet{Iwakami_Nagakura_Yamada:2014b}.       
       We found that there also exists the critical strength of the toroidal magnetic field,
       above which no steady solution exists (right panel in Figure~\ref{fig:Lnu_vs_f}).
\end{enumerate}

We ignore other non-spherical physical processes such as the 
turbulence in progenitors in this paper.
Turbulence in the accretion flow enables explosion by changing the flow property.       
\citet{Murphy_Burrows:2008} performed 1D and 2D dynamical simulations and
suggested that the reduction of critical luminosity is caused by turbulence.
\citet{Murphy_Meakin:2011} examined many turbulent models using Reynolds decomposition
and proposed a global turbulent model that reproduces the profiles and evolution of the simulations 
(see also 2D and 3D dynamical simulations by \citealp{Murphy_Dolence_Burrows:2013}).
Recently, \citet{Mabanta_Murphy:2018} considered 1D spherical
steady models with the turbulence and contended that the turbulent dissipation rather
than the ram pressure from the turbulence reduces the critical luminosity.
The turbulent dissipation is a significant contributor to
successful supernova explosion.
Note that these works ignored the effect of magnetic fields.
\citet{Masada_Takiwaki_Kotake:2015}, on the other hand, 
performed high resolution local simulations with magnetic field
in a 3D thin layer to calculate the magneto-rotational-instability
near the neutrino sphere accurately and found that the convectively
stable layer around the neutrino sphere becomes
fully turbulent due to the MRI. Magnetic field plays an important role on the
turbulence. 

The steady state problems concerning these physical
processes, however, are considered in only 1D spherical models (e.g.~\citealt{Mabanta_Murphy:2018})
and nobody has succeeded in obtaining the 2D steady accretion models with them.
We will study the steady flows with them in 2D self-consistently using our numerical procedure
and analyze the effect from them on the critical neutrino luminosity in the future. 

 \section*{Acknowledgements}

 The authors thank Dr. Kazuya Takahashi and Dr. Wakana Iwakami for helpful discussions. 
 This work was supported by JSPS KAKENHI Grant Numbers 16K17708, 16H03986, 17K18792.
 KF was supported by JSPS Postdoctoral Fellowship for Research Fellowship (16J10223). 
 
\appendix

\section{Rankine-Hugoniot relation on the deformed shock wave}\label{sec:shock}
%

\begin{figure}
 \begin{tabular}{cc}
  \begin{minipage}{0.45\textwidth}
   \psfrag{er}{$\vec{e}_r$}
   \psfrag{vpa}{$\vec{u}_{\parallel}$}
   \psfrag{vpe}{$\vec{u}_{\perp}$}
   \psfrag{chi}{$\chi$}
   \psfrag{rs0}{$r=r_{sph}$}
   \psfrag{rs}{$r=\rs(\theta)$}
   \includegraphics[width=6.5cm]{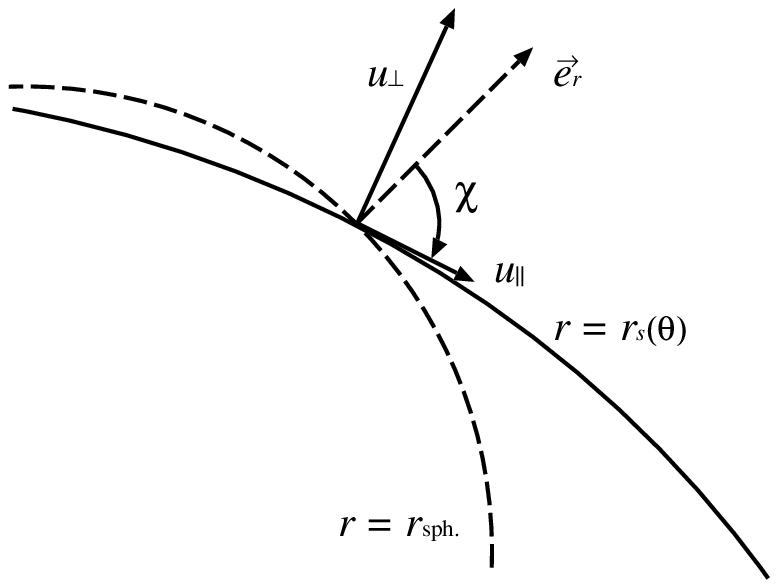}
  \end{minipage}&
   \begin{minipage}{0.45\textwidth}
   \psfrag{x0}{$q=0$}
   \psfrag{x1}{$q=1$}
  \includegraphics[width=8cm,clip]{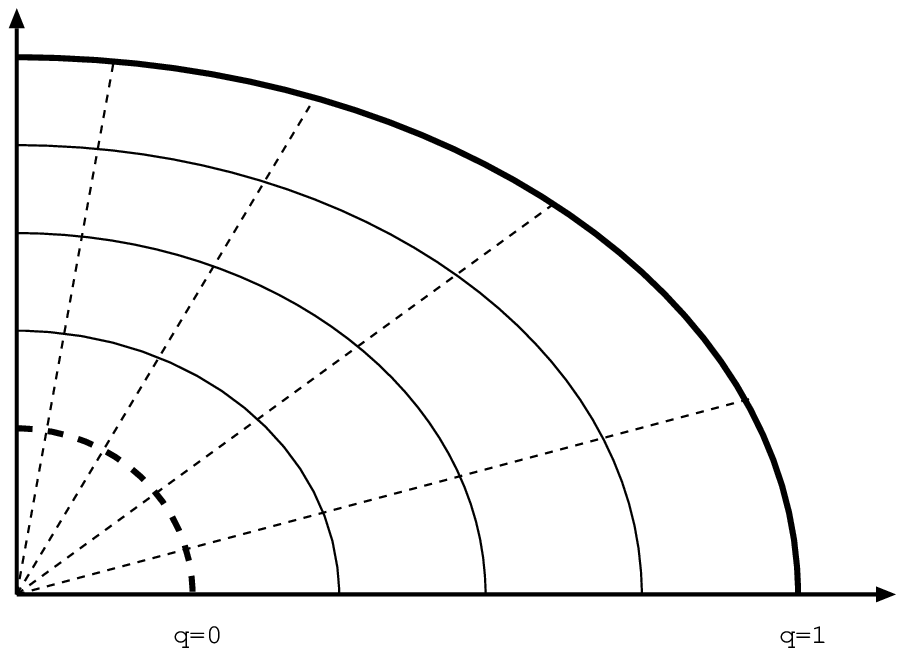}
   \end{minipage}\\
 \end{tabular}
 \caption{
 Schematic pictures of the oblique shock surface (left) and the surface-fitted coordinates (right).
 In the left picture the solid curve denotes
 the shock surface while dashed curve
 represents a sphere with the radius of  $r=r_{\mathrm{sph.}}$.
 The angle between them is defined as $\chi$.
 In the right picture
 $q=0$ corresponds to the neutrino sphere surface whereas $q=1$ is the shock surface.
 }
 \label{fig:coordinates}
\end{figure}

We need to impose the junction condition on the shock surface 
for each variable.
We first define two unit vectors, $u_\perp$ and $u_\parallel$,
each normal and tangential, respectively, to the shock surface given
as $r=r_s(\theta)$. Then $\chi$ is defined to be the angle between the
radial unit vector $e_r$ and $u_\parallel$ as shown
in the left panel of Figure~\ref{fig:coordinates}.
The value of $\chi$ is obtained as follows:
 \begin{subequations}
 \begin{align}
 \vec{u}_{\parallel}\cdot\vec{e}_r =& |\vec{u}_{\parallel}|\cos\chi, \\
  \cos\chi =& \frac{\dif \rs}{\dif\theta}\left\{\rs^2+\left(\frac{\dif \rs}{\dif\theta}\right)^2\right\}^{-\frac{1}{2}}.
  \label{eq:angle_def}
 \end{align}
 \end{subequations}
 It is also useful to adopt the so-called surface-fitted coordinates $(q,\theta')$ defined
 from the spherical coordinates ($r,\theta$) as
\begin{eqnarray}
 \displaystyle q\equiv\frac{r-\rnu}{\rs(\theta)-\rnu},
\end{eqnarray}
\begin{eqnarray}
\theta' = \theta.
\end{eqnarray}
This transformation maps the region between the neutrino sphere and the deformed shock surface 
into a domain given simply as
\begin{eqnarray}
 q \in [0,1], \ \theta' \in [0,\pi].
\end{eqnarray}
See the right panel of Figure~\ref{fig:coordinates}. Noted that this is the domain of
our main concern in this paper.
The derivatives are also transformed as follows:
\begin{subequations} 
\begin{align}
 \frac{\del Q}{\del r} =& \frac{1}{\rs-\rnu}\frac{\del Q}{\del q}, \\
  \frac{\del Q}{\del \theta} =& \frac{\del Q}{\del \theta'}
  -\frac{q}{\rs-\rnu}\frac{\dif\rs}{\dif\theta}\frac{\del Q}{\del q},
\end{align} 
\end{subequations}
where $Q$ is a physical variable such as density or velocity.
Henceforth, we will employ these surface-fitted coordinates alone and
use the same notation $\theta$ instead of $\theta'$
to denote the new angle coordinate.

The MHD Rankine-Hugoniot relation
(e.g. \citealt{Winterhalter:1984}; \citealt{Takahashi_Yamada:2013}) 
 is written as follows:
\begin{subequations}
  \begin{align}
   & \left[ \rho \mvel\cdot\mn\right] = 0,
    \\
   & \left[ \rho \left(\mvel\cdot\mn\right)\mvel
    +\left(p +\frac{B^2}{8\pi}\right)\mn
    -\frac{1}{4\pi}\left(\mB\cdot\mn\right)\mB \right] = 0,
    \\
   & \bigg[ \mvel\cdot\mn\left(\rho\varepsilon +p +\frac{1}{2}\rho u^2 +\frac{B^2}{4\pi}\right)
    \nonumber
    \\
   & -\frac{1}{4\pi}\left(\mB\cdot\mn\right)\left(\mB\cdot\mvel\right)\bigg] = 0,
    \\
   & \left[ \mB\cdot\mn\right] = 0,
    \\
   & \left[ \mn\times\left(\mvel\times\mB\right)\right] = 0,
  \end{align}
  \end{subequations}
  where the background denotes as usual the difference between the upstream 
  and downstream values of the quantity given inside the bracket and 
  we use the ordinary notation $\mn$ for the normal vector to the shock surface.
  One can also express these relations as 
 \begin{subequations}
  \label{eq:Rankine-Hugoniot}
   \begin{align}
    &\rho u_{\perp} = \hat\rho \hat{u}_{\perp},
    \\
    &\rho u_{\perp}\mvel
    +\left(p +\frac{B^2}{8\pi}\right)\mn
    -\frac{1}{4\pi}B_{\perp}\mB
    \nonumber \\
    &=
    \hat\rho \hat{u}_{\perp}\hat{\mvel} +\left(\hat{p} +\frac{\hat{B}^2}{8\pi}\right)\mn
    -\frac{1}{4\pi}\hat{B}_{\perp}\hat{\mB},
    \\
    &u_{\perp}\left(\rho\epsilon +p +\frac{1}{2}\rho u^2 +\frac{B^2}{4\pi}\right)
    -\frac{1}{4\pi}B_{\perp}\left(\mB\cdot\mvel\right)
    \nonumber \\
    &=
    \hat{u}_{\perp}\left(\hat\rho\hat\epsilon +\hat{p} +\frac{1}{2}\hat\rho \hat{u}^2 +\frac{\hat{B}^2}{4\pi}\right)
    -\frac{1}{4\pi}\hat{B}_{\perp}\left(\hat{\mB}\cdot\hat{\mvel}\right),
    \\
    &B_{\perp} = \hat{B}_{\perp},
    \\
    &B_{\perp}\mvel -u_{\perp}\mB
    = \hat{B}_{\perp}\hat{\mvel} -\hat{u}_{\perp}\hat{\mB},
   \end{align}
 \end{subequations}
 where the hat $\, \hat{} \,$ means the upstream value 
 and the subscript $_\perp$ denotes the normal component to the shock surface.  
  We use these relations to obtain the downstream values of physical quantities.
  The rotation and magnetic field upstream the shock are given in Section~\ref{ssec:rotation_law}
  (see also Appendix~\ref{sec:derivation}).

  Since we have assumed a free-falling cold, spherical flow outside the shock surface,
the radial velocity $u_f$, density $\rho_f$ and pressure $p_f$ just ahead of the
shock surface are given as follows:
\begin{subequations}
 \begin{align}
  u_f(r) =& - \sqrt{\frac{2GM}{r}}, \\
 \rho_f (r) =& -\frac{\dot{M}}{4\pi r^2 u_f} = 
  \frac{\dot{M}}{\sqrt{32 \pi^2 GM}} r^{-\frac{3}{2}}, \\
 p_f (r) =& 0.
 \end{align}
 \label{eq:free_fall}
\end{subequations}
Note that the mass accretion rate is positive ($\dot{M}>0$) in this paper.

\section{The rotation law and magnetic-field profile outside the shock wave}
\label{sec:derivation}
%

Here we explain in detail the ideas behind the setting
of the outer boundary conditions for rotation and magnetic field.
If the magnetic field is purely toroidal ($\Br = \Bth = 0$),
there are two conserved quantities: specific angular momentum $\ell$
and  a quantity related with the magnetic torque $\sigma$.
The former comes from the $\varphi$ component of the Euler equation
(Equation \ref{eq:momentum_ph}) and the latter results from the $\varphi$ component of the ideal MHD condition.
They are constant along each streamline. It should be noted that
the specific entropy is not conserved, since heating and cooling are taken into account 
in our formulation unlike the previous works
that assumed the barotropic (adiabatic) condition (e.g.~\citealt{Lovelace_et_al:1986}; \citealt{Fujisawa_et_al:2013}).
The conservations of $\ell$ and $\sigma$ are expressed as
\begin{subequations} 
 \begin{align}
  \ell(\psi) = r \sin \theta \uph, \\
  \sigma(\psi) = \frac{\Bph}{r \sin \theta \rho},
  \label{eq:app_psi}
 \end{align}
\end{subequations} 
where $\ell (\psi)$ and $\sigma(\psi)$ are arbitrary functions of $\psi$,
which is a so-called stream function defined in the following relations:
 \begin{eqnarray}
  \urr \equiv \frac{1}{4 \pi \rho r^2 \sin \theta } \frac{\del \psi}{\del \theta}, \
   \uth \equiv -\frac{1}{4\pi \rho r \sin \theta} \frac{\del \psi}{\del r}.
  \label{eq:app_stream_function}
 \end{eqnarray}
 Since we have assumed that the flow is spherical outside the
 shock surface, the stream function there is given explicitly as
 \begin{eqnarray}
  \psi = \dot{M} \cos \theta.
   \label{eq:psi}
 \end{eqnarray}

We need to specify the functional forms of
 $\ell$ and $\sigma$ to fix the rotation law and the profile of toroidal magnetic field. 
 In this paper following \citet{Yamasaki_Yamada:2005} we specify the angular distributions of
 $\uph$ and $\Bph$ at $r=1000~\mathrm{km}$. 
 Since the stream function depends only on $\theta$ alone (see Equation~\ref{eq:psi}),
 $\ell$ and $\sigma$ are also the functions of $\theta$ alone
 and the $\uph$ and $\Bph$ at $r=1000~\mathrm{km}~(\equiv r_{\mathrm {1000km}})$ are expressed as follows:
  \begin{subequations} 
  \begin{align}
  \uph(r_{1000\mathrm{km}},\theta) &= \frac{\ell(\theta)}{r_{1000\mathrm{km}} \sin \theta}, \\
   \Bph(r_{1000\mathrm{km}},\theta) &=r_{1000\mathrm{km}}\sin \theta \rho_{1000\mathrm{km}} \sigma(\theta)
   \nonumber \\
   &= \frac{\dot{M}}{\sqrt{32\pi^2 G M r_{1000\mathrm{km}}}} \sin \theta \sigma(\theta),
  \end{align}
  \end{subequations}
  where $\rho_{1000 \mathrm{km}}$ is the density at $r_{\mathrm{1000km}}$.
  Just as in \citet{Yamasaki_Yamada:2005} we assume uniform rotation at $r_{\mathrm{1000km}}$
  with a rotational frequency $f_{1000\mathrm{km}}$.
  The $\ell(\theta)$ is given as
  \begin{equation} 
   \ell (\theta) = 2\pi r^2_{1000\mathrm{km}} f_{1000\mathrm{km}} \sin^2 \theta.
     \label{eq:app_rotation_law}
  \end{equation}  
   The azimuthal component of velocity is also obtained as follows:
  \begin{equation} 
   \uph(r, \theta) = 2\pi \frac{r_{1000\mathrm{km}}^2}{r} \sin \theta f_{1000\mathrm{km}}.
  \end{equation}
  As for $\sigma$, we simply assume that it is independent of $\theta$ at $r_{\mathrm{1000km}}$:
      \begin{equation} 
	\sigma(\theta)
	= -\frac{\sqrt{32\pi^2 G M}}{\dot{M}} B_{1000\mathrm{km}} \sqrt{r_{1000\mathrm{km}}}.
      \end{equation}
      Then the profile of the toroidal magnetic field is given as follows:
      \begin{equation} 
       \Bph (r, \theta) = B_{1000\mathrm{km}} \sqrt{\frac{r_{1000\mathrm{km}}}{r}} \sin \theta,
       \end{equation}
   where the $B_{1000\mathrm{km}}$ is the strength of toroidal magnetic field at $r_{\mathrm{1000km}}$.
   Note that the profiles of $\uph$ and $\Bph$ thus obtained apparently
   satisfy the boundary conditions on the symmetry axis and the equator 
   in Equation~(\ref{eq:boundary_condition}).
   
   Next we consider the case, in which the magnetic field has both 
   poloidal and toroidal components. Then
   the situation is more complicated with $\ell$ and $\sigma$ being
   no longer independent of each other (\citealt{Fujisawa_et_al:2013}).
   From the continuity equation of magnetic field
   ($\nabla \cdot \mB = 0$), the magnetic flux function $\Psi$ is defined
   to give the poloidal magnetic-field components as
 \begin{eqnarray}
  \Br   = \frac{1}{r^2 \sin \theta}\frac{\del \Psi}{\del \theta}, \,\,\,\,
  \Bth  =-\frac{1}{r   \sin \theta}\frac{\del \Psi}{\del r}.
 \end{eqnarray}
 Since the poloidal magnetic-field lines are coincide with the streamlines
 in ideal MHD, the magnetic flux function is also constant along each streamline,
 \begin{eqnarray}
  \Psi \equiv \Psi(\psi).
 \end{eqnarray}
 Then $\ell$ and $\sigma$ are expressed in terms of the magnetic flux function,
 which is in turn a function of the stream function alone (see \citealt{Fujisawa_et_al:2013}) as
      \begin{subequations} 
	\begin{align}
	 \label{eq:app_ell_sigma2_a}
	\ell(\psi) &= r \sin \theta \uph - r \sin \theta \frac{d \Psi(\psi)}{d \psi} \Bph, \\
	 \label{eq:app_ell_sigma2_b}
	 \sigma(\psi) &= \frac{\Bph}{r \sin \theta \rho} - \frac{4\pi}{r \sin \theta} \uph \frac{d \Psi (\psi)}{d \psi}.
	\end{align}
         \label{eq:app_ell_sigma2}
      \end{subequations}      
      Note that the specific angular momentum ($r\sin\theta \uph$)
      is no longer conserved along a streamline
      but the $\ell$ is sill and $\sigma$ too remains a conserved quantity although it also
      contains a term that depends on $\Psi$.

      Just as in the previous case, in which the purely toroidal magnetic field is considered,
      we set the functional form of $\Psi$
      outside the shock surface from the condition that the  streamlines and
      hence the poloidal magnetic-field line are spherical
      as follows:
      \begin{subequations} 
   \begin{align}
    \Psi &= -r_{\mathrm{1000km}}^2 B_0 \cos \theta, \\    
      \label{eq:app_poloidal}
    \Br(r,\theta) &= \frac{r^2_{\mathrm{1000km}}}{r^2} B_0, \\
    \frac{d\Psi}{d\psi} &= -r_{\mathrm{1000km}}^2 \frac{B_0}{\dot{M}} ={\mathrm{constant}},
   \end{align}
      \end{subequations}
 where $B_0$ is a constant corresponding to the strength of poloidal magnetic field at $r_\mathrm{1000km}$.
 It is apparent from Equation~(\ref{eq:app_poloidal}) that the radial 
 magnetic field does not depend on $\theta$ outside the shock surface.
 Since the $\theta$ component of magnetic field vanishes $\Bth = 0$ outside the shock surface,
 the $\varphi$ component of the ideal MHD condition given in Equation \eqref{eq:basics_mag} becomes
 \begin{subequations}
  \begin{align}
    \label{eq:mag_field_ph_2}
  &\frac{\partial}{\partial r}\left(r\Br\uph -r\Bph\urr\right) = 0
  \\ 
   &\Rightarrow \left(r\Br\uph -r\Bph\urr\right) = C_1(\theta), \\
   &\Rightarrow \frac{\uph}{r \sin \theta} \frac{d\Psi}{d\psi} \frac{\partial \psi}{\partial \theta}
   - \frac{\Bph}{4\pi \rho r \sin \theta} \frac{\partial \psi}{\partial \theta} = C_1(\theta).
  \end{align}
 \end{subequations}
 In the above equations, $C_1$ is a function of $\theta$ alone and is related with $\sigma$
 as follows (see Equation~\ref{eq:app_ell_sigma2_b}): 
\begin{eqnarray}
 \sigma(\theta) = - 4\pi \left( \frac{\partial \psi}{\partial \theta}  \right)^{-1} C_1 (\theta).
\end{eqnarray}
The $\varphi$ component of the equations of motion (Equation~\ref{eq:momentum_ph})
can be recast in a similar way as
\begin{subequations}
 \begin{align}
   \label{eq:mhd_flow}
& \urr \frac{\partial\uph}{\partial r} + \urr \frac{\uph}{r}
 -\frac{\Br}{4\pi\rho}\left( \frac{\partial\Bph}{\partial r} +\frac{\Bph}{r} \right)=0
 \\
 &\Rightarrow \frac{\partial}{\partial r}(r\uph)
  - \frac{B_r}{4\pi \rho \urr} \frac{\partial}{\partial r} \left(r \Bph \right)
  = 0
 \\
 &\Rightarrow
 r\uph - \frac{B_r}{4\pi \rho \urr} \left(r \Bph \right) = C_2(\theta), \\
 &\Rightarrow
 r\uph - r \frac{d\Psi}{d\psi} B_\varphi = C_2(\theta),
 \end{align}
\end{subequations}
where we use the fact that $\Br / (\rho \urr)$ is constant in the
 current case.  $C_2$ is again related with $\ell$ (see Equation~\ref{eq:app_ell_sigma2_a}) as
\begin{eqnarray}
 \ell(\theta) = \sin \theta C_2(\theta).
\end{eqnarray}
Using $C_1$ and $C_2$ instead of $\ell$ and $\sigma$,
we express $\Bph$ and $\uph$ as
\begin{subequations} 
 \begin{align}
 \Bph &= \left(r \frac{\urr}{\Br} - \frac{r^3 \Br}{\dot{M}} \right)^{-1}  \left(C_2 - \frac{C_1}{\Br}\right),
  \\
 \uph &= \left(r - \frac{r^3 \Br^2}{\dot{M}\urr}\right)^{-1} \left(C_2 - \frac{r^2 \Br}{\urr \dot{M}} C_1 \right).
 \end{align}
\end{subequations}
It is apparent from these expressions that there may exist a singularity on a surface,
where the radial velocity becomes equal to the Alfv\'en velocity $u_A^2 = B_r^2 / (4\pi \rho)$.
In this paper, we do not deal with this problem but simply avoid it by choosing 
the value of $B_0$ so that no such singular surface would not be encountered. 

In actual computations, we set the values of $f_{\mathrm{1000km}}$ and $B_{\mathrm {1000km}}$ and
obtain those of $\uph$ and $\Bph$ at $r_\mathrm{1000km}$ as
\begin{subequations} 
 \begin{align}
 \uph(r_{\mathrm{1000km}}, \theta) &= 2\pi r_{1000\mathrm{km}} \sin \theta f_{1000\mathrm{km}},
  \\
 \Bph (r_{\mathrm{1000km}}, \theta) &= B_{1000\mathrm{km}} \sin \theta,
 \end{align}
\end{subequations}
and then we derive the values of $\uph(r_s, \theta)$ and $\Bph(r_s,\theta)$
just ahead of the shock wave by solving Equations~(\ref{eq:mag_field_ph_2}) and (\ref{eq:mhd_flow}) numerically.

 \section{Numerical procedure}\label{sec:flowchart}

Here we outline the procedure to solve the non-linear differential equations
that describe the steady, shocked, non-spherical accretion flows. 
The system of equations given in Section~\ref{sec:basic} is formally written as
\begin{eqnarray}
 \label{eq:basics0}
  \mathcal{A}\left(\bm{Q}\right)\frac{\partial\bm{Q}}{\partial q}
  +\mathcal{B}\left(\bm{Q}\right)\frac{\partial\bm{Q}}{\partial \theta} +\mathcal{C}\left(\bm{Q}\right)
  = 0,
\end{eqnarray}
where we introduce the variable vector~$\bm{Q}\equiv\left[\rho\ \urr\ \uth\ \uph\ T\ \Br\ \Bth\ \Bph\right]^{T}$
for brevity; $\mathcal{A}\left(\bm{Q}\right), \mathcal{B}\left(\bm{Q}\right)$
and $\mathcal{C}\left(\bm{Q}\right)$ are matrix-valued nonlinear function of $\bm{Q}$.
Note that the shock surface corresponds to $q=1$ on the surface-fitted coordinates. 
Equations~\eqref{eq:basics0} are discretized at the cell-center in the $q$-mesh
while at the grid point in the $\theta$-mesh as follows:
\begin{eqnarray}
 \bm{F}_{j-1,k}\equiv
 \mathcal{A}\left(\bm{Q}_{j-\frac{1}{2},k}\right)\frac{\bm{Q}_{j,k}-\bm{Q}_{j-1,k}}{q_{j}-q_{j-1}}
+\mathcal{B}\left(\bm{Q}_{j-\frac{1}{2},k}\right)\frac{\bm{Q}_{j-\frac{1}{2},k+1}-\bm{Q}_{j-\frac{1}{2},k-1}}{2\Delta \theta}
  +\mathcal{C}\left(\bm{Q}_{j-\frac{1}{2},k}\right)
  = 0, 
  \label{eq:procedure_eqs}
\end{eqnarray}
where $\bm{Q}_{j-\frac{1}{2},k}$ is defined as
\begin{eqnarray}
  \bm{Q}_{j-\frac{1}{2},k} = \frac{1}{2}\left(\bm{Q}_{j-1,k} + \bm{Q}_{j,k} \right).
  \label{eq:procedure_eqs2}
\end{eqnarray}
The set of algebraic nonlinear Equations~\eqref{eq:procedure_eqs} is solved
numerically with the W4 method in Appendix~\ref{sec:W4}.

We take the following steps to obtain a solution:
\begin{enumerate}
 \item We first fix the five model parameters $\dot{M}$, $L_\nu$, $f_{1000\mathrm{km}}$, $B_{1000 \mathrm{km}}$ and $B_0$.
       Then the inner boundary $r_{\nu}$ is determined from Equation \eqref{eq:neutrino_sphere}.
       We give an initial guess to the shape of shock surface $r_s(\theta)$,
       which determines the angle $\chi$ in Equation \eqref{eq:angle_def}.

 \item The density, velocity and pressure outside the shock are given in
       Equation (\ref{eq:free_fall}) and other variables such as $\uph$, $\Bph$ and $\Br$ are determined
       as explained in Appendix~\ref{sec:derivation}.
       
 \item Given the upstream quantities, the corresponding 
       downstream quantities are obtained as mentioned
       by solving the Rankine-Hugoniot relation given in Equation~(\ref{eq:Rankine-Hugoniot}).
       In particular, the radial and $\theta$ components of the velocity and magnetic field  just
       behind the shock front are given in terms of the 
       parallel ($\upar, \Bpar$) and perpendicular ($\uper, \Bper$)
       components of velocity and magnetic field as
       \begin{subequations}
	\begin{align}
	 \urr =& \upar \cos\chi +\uper\sin\chi,\\
	 \uth =& \upar \sin\chi -\uper\cos\chi,\\
	 \Br  =& \Bpar \cos\chi +\Bper\sin\chi,\\
	 \Bth =& \Bpar \sin\chi -\Bper\cos\chi.	 
	\end{align}
       \end{subequations}

 \item We integrate the basic equations (Equations~\ref{eq:basics_hydro} and ~\ref{eq:basics_mag})
       inward from the outer boundary~$q=1$ to the inner boundary~$q=0$.
       If the density $\rho$ obtained at the inner boundary differs from
       the specified value
       $\rho =10^{11}~\mathrm{g~cm^{-3}}$,
       we modify the shock surface, changing $r_{\mathrm{sph}}$ and/or $\epsilon$
       and repeat the iteration steps 3 and 4 until the correct value of $\rho$ is
       obtained at the inner boundary. 
\end{enumerate}
We change the value of neutrino luminosity $L_{\nu}$ and repeat the
above procedures until we obtain a series of solutions up to 
the critical point.

 \section{Details of the W4 method}\label{sec:W4}

The Newton-Raphson method is one of the simplest and the most commonly used methods
to solve systems of nonlinear equations numerically. It is very efficient,
converging to a root quadratically, if it really converges,
which happens normally when the initial guess 
is sufficiently close to the root (\citealt{Numerical_Recipes}).
The W4 method of our own devising is an iterative relaxation method just like the
Newton-Raphson method and some of its extensions, 
quasi-Newton method (e.g. Broyden's method, \citealt{Broyden:1965}).
Unlike these Newton-type methods, the W4 method uses acceleration and damping terms
in addition to the velocity term for convergence. In fact, thanks to these terms, the 
W4 method shows better global convergence: we obtain the solutions that Newton-type cannot. 
We explain the W4 method in detail, staring
with a brief of introducing its new extension.

\subsection{Newton-Raphson method}

In this paper we solve a system of nonlinear equations numerically to obtain steady
accretion flows through a standing shock wave onto a PNS. 
Such a problem can be written generically as
\begin{eqnarray}
 F_i (x_1, x_2, \cdots, x_N) = 0 \hspace{10pt} i = 1,2, \cdots, N,
  \label{eq:F_i}
\end{eqnarray}
for $N$ variables $x_i,i=1,2,\cdots,N$. For notational convenience we
let $\bm{x}$ and $\bm{F}$. In solving this type of equations, the 
Newton-Raphson method is probably the first choice.
In the Newton-Raphson method, each function $F_i$ is Taylor-expanded to the
first order at a certain $x$ (\citealt{Numerical_Recipes}) as
\begin{eqnarray}
 F_i(\bm{x} + \delta \bm{x}) = F_i(\bm{x}) + \sum_{j=1}^{N} \frac{\del F_i}{\del x_j} \delta x_j +
{\cal O} (\delta \bm{x}^2).
\end{eqnarray}
The Jacobian matrix is then introduced 
\begin{eqnarray}
 J_{ij} \equiv \frac{\del F_i}{\del x_j}.
\end{eqnarray}
We solve the linearized equations 
\begin{eqnarray}
 F_i(\bm{x} + \delta \bm{x}) = F_i(\bm{x}) + \sum_{j=1}^{N} J_{ij} \delta x_j = 0,
\end{eqnarray}
for $\delta x_j$ as
\begin{eqnarray}
 \label{eq:NR1}
 \delta x_i = - \sum_{j=1}^N J^{-1}_{ij} F_j,
\end{eqnarray}
where $J^{-1}_{ij}$ is the inverse of the Jacobian matrix.
The value of $x$ is then incremented by $\delta x$ and Equation~(\ref{eq:F_i}) is solved for this
new $x$. The procedure is repeated until the sequence $x_n$ determined from 
\begin{eqnarray}
  \label{eq:NR2}
 \bm{x}^{n+1} = \bm{x}^{n} + \delta \bm{x},
\end{eqnarray}
is converged. 
Whereas in the Newton-Raphson method the correction terms $\delta \bm{x}$ is calculated
with the inverse of the Jacobian matrix as in Equation (\ref{eq:NR1}),
it is obtained in different ways (e.g. good and bad Broyden's method, \citealt{Broyden:1965})
in the secant methods or other Newton-type methods.
Regardless, it is important that all these methods employ Equation~(\ref{eq:NR2}) for 
a single step evolution.
We modify this in the W4 method.

\subsection{W4 method}

Equation~(\ref{eq:NR2}), which is used for the single-step evolution commonly in the Newton-type methods,
may be recast in to the following suggestive form:
If we introduce a time step width $d\tau$,  Equation (\ref{eq:NR2}) becomes as
\begin{eqnarray}
 \frac{\bm{x}^{n+1} - \bm{x}^n}{\Delta\tau} = \bm{f}(\bm{x}), \hspace{10pt}
  \bm{f}(\bm{x}) \equiv \frac{\delta \bm{x}}{\Delta\tau}
  = - \frac{\mathbf{J}^{-1} \bm{F}}{\Delta \tau}.
\end{eqnarray}
where $\Delta\tau$ is a fictitious time step, which is arbitrary for the moment.

Then the resultant equation may be interpreted as an approximation to the
following first-order differential equations:
\begin{eqnarray}
 \frac{d\bm{x}}{d\tau} = \bm{f}(\bm{x}).
\end{eqnarray}
It is probably not surprising then to consider differential equations of second order
instead of first order given as follows:
\begin{eqnarray}
 \frac{d^2\bm{x}}{d\tau^2}  + \mathbf{M}_1 \frac{d\bm{x}}{d\tau} + \mathbf{M}_2 \bm{F} = 0,
\end{eqnarray}
where $\mathbf{M}_1$ and $\mathbf{M}_2$ are matrices somehow related with the Jacobian matrix.
Note that these equations remind us of the forced oscillation of connected springs  with damping.
This analogy is important and useful in considering the behaviour of solutions. Introducing 
"momentum" $\bm{p}$, we can decompose these second order differential equations
into two sets of first order differential equations as follows:
\begin{eqnarray}
  \label{eq:W4_1}
  \frac{d\bm{x}}{d\tau} = \mathbf{X} \bm{p}, \hspace{10pt}
 \frac{d\bm{p}}{d\tau} = - 2\bm{p} - \mathbf{Y} \bm{F},
\end{eqnarray}
where $\mathbf{X}$ and $\mathbf{Y}$ are matrices related with $\bm{F}$ and the Jacobian matrix.
Finite-differencing these first-order differential equations, we obtain a new recurrence formula for the
single-step evolution as
\begin{eqnarray}
\label{eq:W4_2}
 \bm{x}^{n+1} = \bm{x}^{n} + \Delta \tau \mathbf{X} \bm{p}^n, \hspace{10pt}
 \bm{p}^{n+1} = (1-2\Delta \tau) \bm{p}^{n} - \Delta \tau \mathbf{Y} \bm{F}.
\end{eqnarray}
These are the single-step evolution equations adopted in the W4 method.
It should be apparent that this is an extension of Equation (\ref{eq:NR2}). 
It should be emphasized, however, that there is a greater degree of freedom 
in choosing $\mathbf{X}$ and $\mathbf{Y}$. Indeed we have studied
various choices and found that 
UL decomposition of the Jacobian matrix show a nice performance in finding roots.
In this decomposition $\mathbf{J} \equiv \mathbf{U}\mathbf{L}$, $\mathbf{U}$ and $\mathbf{L}$
are upper and lower triangular matrices given, respectively, as follows:
\begin{eqnarray}
 \mathbf{U} = \left(
      \begin{array}{cccc}
       u_{11} & u_{12} &  u_{13}   \\
           0  & u_{22} &  u_{23}   \\
           0  &   0    &  u_{33}   \\
      \end{array}
	     \right),\hspace{10pt}
 \mathbf{L} = \left(
      \begin{array}{cccc}
               1  &   0            &  0                       \\
       \ell_{21}  &   1            &  0                       \\
       \ell_{31}  &   \ell_{32}    &  1                       \\
      \end{array}
	     \right), 
\end{eqnarray}
where $u_{ij}$ and $\ell_{ij}$ are elements of these matrices for the case of a $3\times 3$ matrix.
Setting $\mathbf{X} = \mathbf{L}^{-1}$ and $\mathbf{Y} = \mathbf{D}^{-1}$ as well as $\Delta \tau = 0.5$,
we obtain in the version we refer to as the UL-W4 method (\citealt{W4})
\begin{eqnarray}
 \bm{x}^{n+1} = \bm{x}^{n} + \frac{1}{2} \mathbf{L}^{-1}_n        \bm{p}^{n}, \hspace{10pt}
 \bm{p}^{n+1} =-\frac{1}{2} \mathbf{U}^{-1}_n\bm{F}(\bm{x}^n), \hspace{10pt}
\label{eq:ULW4}
\end{eqnarray}
where $\mathbf{L}_{n}^{-1}$ and $\mathbf{U}_{n}^{-1}$
denote $\mathbf{L}^{-1}$ and $\mathbf{U}^{-1}$
at $n$-th time step, respectively.
Note that combined Equations~(\ref{eq:ULW4})
\begin{eqnarray}
 \bm{x}^{n+1} = \bm{x}^{n} - \frac{1}{4} \mathbf{L}^{-1}_n \mathbf{U}^{-1}_{n-1} \bm{F}(\bm{x}^{n-1}), 
\end{eqnarray}
differ from the single-step evolution equation in the Newton-Raphson method because
$\mathbf{L}^{-1}_{n} \mathbf{U}^{-1}_{n-1} \neq \mathbf{J}^{-1}_{n}$ and
$\mathbf{L}^{-1}_{n} \mathbf{U}^{-1}_{n-1} \neq \mathbf{J}^{-1}_{n-1}$ in general.
It should be also stressed that 
the product tends to the inverse of the Jacobian matrix 
$\mathbf{L}^{-1}_{n} \mathbf{U}^{-1}_{n-1} \simeq \mathbf{J}^{-1}_{n-1} \simeq \mathbf{J}^{-1}_{n}$
as the sequence approaches the root. These properties turn out to be crucial for 
both global and local convergences (\citealt{W4}). 

Another useful choice of $\mathbf{X}$ and $\mathbf{Y}$ is derived from
the LH decomposition of the Jacobian matrix, which is in turn based on QR decomposition of
the transposed Jacobian matrix~$(\mathbf{J}^{T} \equiv \mathbf{Q}\mathbf{R})$,
where $\mathbf{H}$ is a Householder matrix, $\mathbf{Q}$ is an
orthogonal matrix and $\mathbf{R}$ is an upper triangular matrix.
Let us describe the procedure of these decompositions for 
a $3\times 3$ Jacobian matrix, which is expressed as 
\begin{eqnarray}
 \mathbf{J}^{T} = \mathbf{A}_{(0)}
  \equiv 
  \left(
   \begin{array}{ccc}
    a_{11}^{(0)} & a_{12}^{(0)} &  a_{13}^{(0)} \\
    a_{21}^{(0)} & a_{22}^{(0)} &  a_{23}^{(0)} \\
    a_{31}^{(0)} & a_{32}^{(0)} &  a_{33}^{(0)}
   \end{array}
  \right).
\end{eqnarray}
First, we extract the first column of this matrix as a column 
vector~$\bm{a}_{(0)}\equiv \left[a_{11}^{(0)}\ a_{21}^{(0)}\ a_{31}^{(0)}\right]^{T}$,
and make another vector~$\bm{b}_{(0)}$ with the norm $|\bm{a}_{(0)}|$ as
\begin{eqnarray}
 \bm{b}_{(0)} \equiv \left[ -{\rm sign}\left(a_{11}^{(0)}\right)|\bm{a}_{(0)}|  \ 0\ 0\right]^{T},
\end{eqnarray}
where we use the sign function. 
The Householder matrix~$\mathbf{H}_{(0)}\equiv \mathbf{E} -2\bm{c}_{(0)}\bm{c}_{(0)}^{T}$,
where $\mathbf{E}$ denotes the identity matrix,
 \begin{eqnarray}
 \bm{c}_{(0)} \equiv \frac{\bm{a}_{(0)} -\bm{b}_{(0)}}{|\bm{a}_{(0)} -\bm{b}_{(0)}|},
\end{eqnarray}
and transforms the vector~$\bm{a}_{(0)}$ into $\bm{b}_{(0)}$ or vice versa: 
$\mathbf{H}_{(0)}\bm{a}_{(0)}=\bm{b}_{(0)}$ and $\mathbf{H}_{(0)}\bm{b}_{(0)}=\bm{a}_{(0)}$.
Multiplying this Householder matrix with $\mathbf{A}_{(0)}$, we obtain 
\begin{eqnarray}
 \mathbf{H}_{(0)} \mathbf{A}_{(0)}
  \equiv \mathbf{A}_{(1)} =
  \left(
   \begin{array}{ccc}
    r_{11} & r_{12} &  r_{13} \\
    0 & a_{22}^{(1)} &  a_{23}^{(1)} \\
    0 & a_{32}^{(1)} &  a_{33}^{(1)}
   \end{array}
  \right),
\end{eqnarray}
where $r_{1j}$ and $a_{ij}^{(1)}$ are the components of the matrix $\mathbf{A}_{(1)}$ thus
obtained, which are non-vanishing in general.

We repeat the same process to the lower right $2\times 2$ submatrix.
We again extract the column vector~$\bm{a}_{(1)}\equiv \left[a_{22}^{(1)}\ a_{32}^{(1)}\right]^{T}$
and make a new vector~$\bm{b}_{(1)}\equiv \left[ -{\rm sign}\left(a_{22}^{(1)}\right)|\bm{a}_{(1)}| \ 0\right]^{T}$
and construct another Householder matrix
\begin{eqnarray}
 \mathbf{H}_{(1)} \equiv 
  \left(
   \begin{array}{cc}
    1 & \bm{0}  \\
    \bm{0} &\quad  \mathbf{E} -2\bm{c}_{(1)}\bm{c}_{(1)}^{T}
   \end{array}
  \right),
\end{eqnarray}
where $\mathbf{E}$ and $\bm{c}_{(1)}\equiv (\bm{a}_{(1)} -\bm{b}_{(1)}) / |\bm{a}_{(1)} - \bm{b}_{(1)}|$
are now $2\times 2$ matrices. 
We multiply it with $\mathbf{A}_{(1)}$ and obtain the following:
\begin{eqnarray}
 \mathbf{H}_{(1)} \mathbf{A}_{(1)}
  \equiv \mathbf{A}_{(2)} =
  \mathbf{H}_{(1)} \mathbf{H}_{(0)} \mathbf{J}^{T}
  =
  \left(
   \begin{array}{ccc}
    r_{11} & r_{12} &  r_{13} \\
    0 & r_{22}       &  r_{23}  \\
    0 &    0          &  r_{33}
   \end{array}
  \right) = \mathbf{R}.
\end{eqnarray}
It is clear from these equations that the product of the two 
Householder matrices make the transposed Jacobian matrix to the upper
triangular matrix $\mathbf{R}$. Since the Householder matrix $\mathbf{H}$ is orthogonal and
symmetric ($\mathbf{H}^{T} = \mathbf{H}^{-1} = \mathbf{H}$) by construction, 
the Jacobian matrix is decomposed as follows:
\begin{eqnarray}
 \mathbf{J}^T = \mathbf{H}_{(0)} \mathbf{H}_{(1)} \mathbf{R} \ \Rightarrow \
 \mathbf{J} = (\mathbf{H}_{(0)} \mathbf{H}_{(1)} \mathbf{R} )^T  = \mathbf{L} \mathbf{H}_{(1)} \mathbf{H}_{(0)},
\end{eqnarray}
where $\mathbf{L}$ is the transpose of $\mathbf{R}$ and a lower triangular matrix.

Finally, we set $\mathbf{X}$ and $\mathbf{Y}$ as $\mathbf{X}= \mathbf{H}_{(1)}\mathbf{L}^{-1}$
and $\mathbf{Y} = \mathbf{H}_{(0)}$.
Then, Equations~(\ref{eq:W4_2}) become
\begin{eqnarray}
 \bm{x}^{n+1} = \bm{x} + \frac{1}{2} \mathbf{H}_{(0)}        \bm{p}^{n}, \hspace{10pt}
 \bm{p}^{n+1} =-\frac{1}{2} \mathbf{H}_{(1)} \mathbf{L}^{-1}\bm{F}(\bm{x}). \hspace{10pt}
\end{eqnarray}
Although we explain the procedure for the $3\times 3$ Jacobian matrix, the generalization to
other dimensions should be obvious: if the dimension is larger than 4, we set $\mathbf{X}=\mathbf{H}_{(0)}$
and $\mathbf{Y} = \cdots \mathbf{H}_{(3)} \mathbf{H}_{(2)} \mathbf{H}_{(1)} \mathbf{L}^{-1}$.

We have applied the W4 method to various problems and found more often not that it can 
find a root even initial guesses are not close to the root and other
 Newton-type methods such as the original Newton-Raphson method and Broyden method
 fail to reach it. This is also the case for the problem of our interest in this paper, that is,
 solving the (M)HD equations numerically to obtain
 the steady accretion flows through a stalled shock wave onto the PNS.
 These facts  indicate that the W4 method of our own devising has a better global convergence
 property than those other methods. We stress that the UL or LH decomposition introduced above
 are just two useful possibilities and the W4 method has
 much greater possibilities. We refer readers to \citet{W4} for more mathematical aspects
 of the method and numerical tests.

 \section{Convergence test}\label{sec:convergence}

 \begin{figure}[h]
  \begin{tabular}{cc}
   \includegraphics[width=8.5cm,clip]{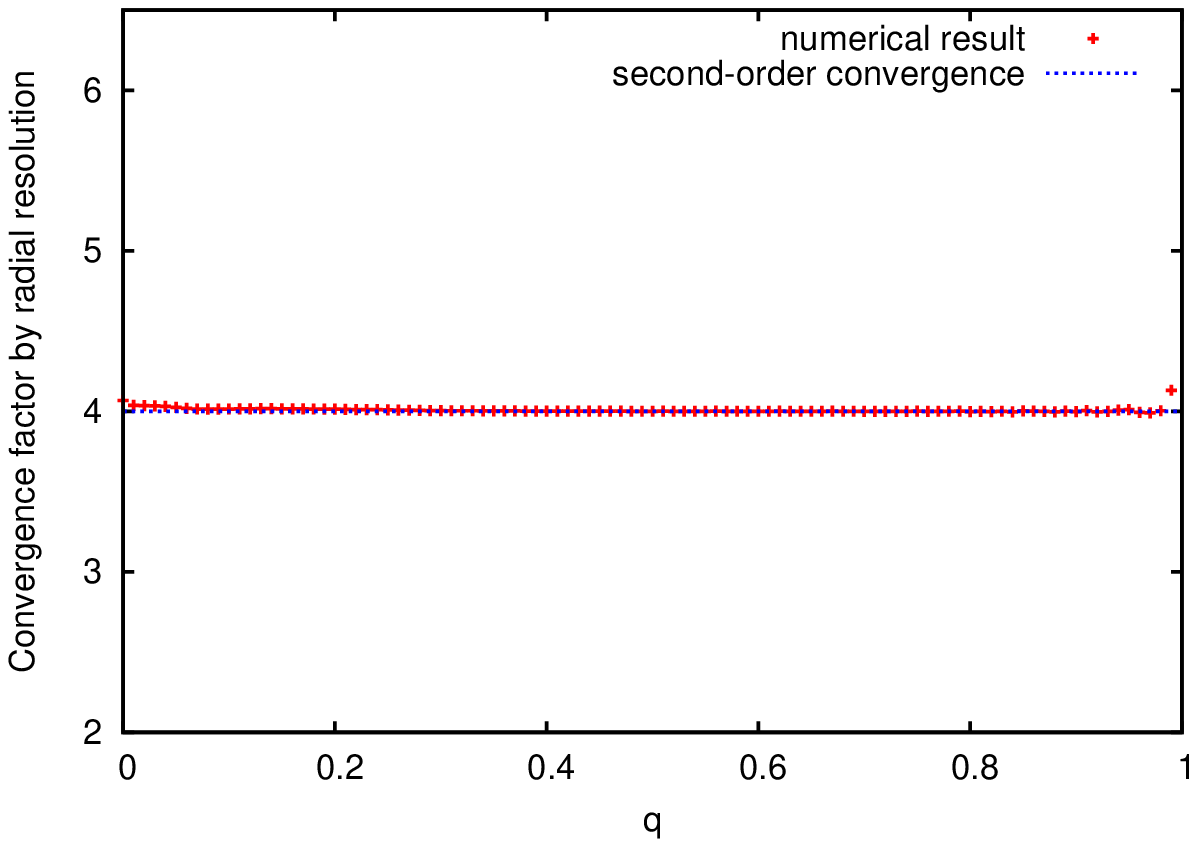} &
   \includegraphics[width=8.5cm,clip]{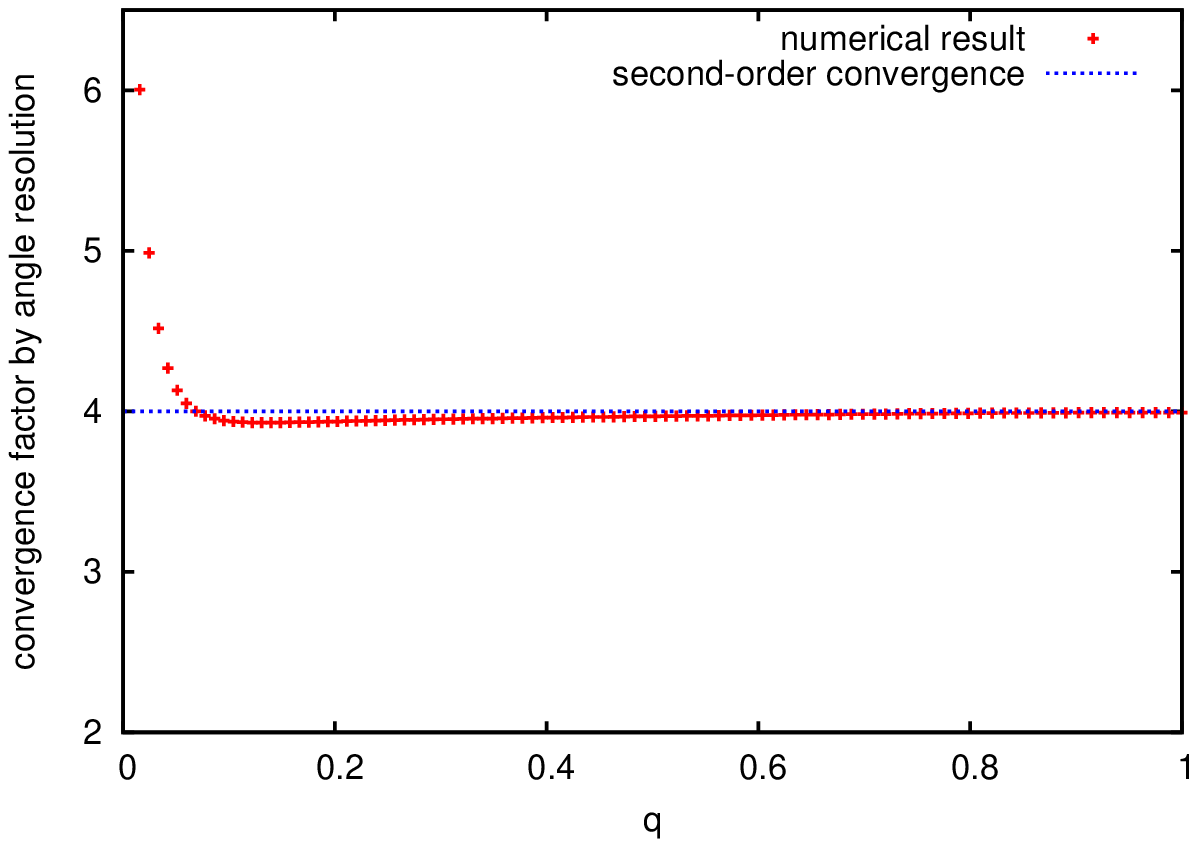}
  \end{tabular}
  \caption{
  Convergence factors $Q_q$ (left) and $Q_\theta$ (right)
  as functions of the $q$ with $N_{q} =200$ and $N_\theta = 20$.
  The mass accretion rate and neutrino luminosity are set to
  $\dot{M} = \mathrm{2.0~M_{\odot}~s^{-1}}$ and $L_{\nu}=4.5\times 10^{52}~\mathrm{ergs~s^{-1}}$, respectively.
  The rotational frequency is $f_{\mathrm{1000km}}=0.03~\mathrm{s^{-1}}$
  and the strengths of the toroidal and poloidal magnetic fields are given as  
  $B_{\mathrm{0}}=10^{6}~\mathrm{G}$ and $B_{\mathrm{1000km}}=10^{6}~\mathrm{G}$ at the outer boundary.
  The dotted blue line denotes the second-order convergence ($Q = 4$).
  }
  \label{fig:ConvergenceTest}
 \end{figure}

We check the convergence of the numerical solutions we obtain in this paper by changing
the number of grid points. We use the results with three different numbers of grid points and
check the convergence factors $Q_q$ and $Q_\theta$ (e.g. \citealt{Okawa_et_al:2014}) as follows:
\begin{subequations}
 \begin{eqnarray}
   Q_q      \equiv \left|\frac{\phi_{2N_q}-\phi_{N_q}}{\phi_{N_q}-\phi_{N_q/2}} \right|, \\
  Q_\theta  \equiv  \left|\frac{\phi_{2N_\theta}-\phi_{N_\theta}}{\phi_{N_\theta}-\phi_{N_\theta/2}} \right|,
 \end{eqnarray}
   \label{eq:ConvergenceFactor}
\end{subequations}
where $N_q$ and $N_\theta$ denote the numbers of grid in the $q$-direction
and $\theta$-direction and $\phi_{N_q}$ and $\phi_{N_\theta}$ are
physical quantities calculated with $N_r$-grids and $N_\theta$-grids.
We here use the $\theta$-averaged tangential component of velocity 
\begin{eqnarray}
 \phi(q) = \frac{1}{N_\theta}\sum_{k=1}^{N_\theta}\uth(q,\theta_k),
\end{eqnarray}
because it is easily influenced by the numerical errors at boundaries.
Since we use the cell-centered second-order discretization in both directions as
in Equations~(\ref{eq:procedure_eqs} and \ref{eq:procedure_eqs2}), 
the convergence factors must be $Q\sim 4$ if the numerical results are converged well.
In the convergence test the mass accretion rate and neutrino luminosity are set to
$\dot{M} = \mathrm{2.0~M_{\odot}~s^{-1}}$ and $L_{\nu}=4.5\times 10^{52}~\mathrm{ergs~s^{-1}}$, respectively.
The rotational frequency is $f_{\mathrm{1000km}}=0.03~\mathrm{s^{-1}}$
and the strengths of the toroidal and poloidal magnetic fields are given as  
$B_{\mathrm{0}}=10^{6}~\mathrm{G}$ and $B_{\mathrm{1000km}}=10^{6}~\mathrm{G}$ at the outer boundary.

Figure~\ref{fig:ConvergenceTest} displays the convergence factor $Q_q$ (left) and $Q_\theta$ (right)
as functions of $q$ with $N_q = 200$ and $N_\theta=20$.
It is obvious that our code actually has the second-order convergence with respect to
both $N_q$ and $N_\theta$ with these grid numbers.
Note that the solutions are almost unchanged with $N_\theta \sim 20$, especially near the neutrino sphere.
The actual number $N_q$, on the other hand, depends on the mass accretion rate $\dot{M}$ and
the numerical scheme employed but the typical value is $N_q \sim 100$ for high
mass accretion rates and $N_q \sim 1000$ for models with low mass accretion rates.
The LH-W4 method requires smaller values of $N_q$ than the UL-W4 method. 
We hence set $N_\theta \sim 20$ and $N_q \sim 100$ for models with high mass accretion rates
and $N_q \sim 1000$ for models with low mass accretion rates
for the numerical computations in this paper.





\bibliographystyle{aasjournal.bst}

\end{document}